\input harvmac
\input epsf
%
\newbox\hdbox%
\newcount\hdrows%
\newcount\multispancount%
\newcount\ncase%
\newcount\ncols
\newcount\nrows%
\newcount\nspan%
\newcount\ntemp%
\newdimen\hdsize%
\newdimen\newhdsize%
\newdimen\parasize%
\newdimen\spreadwidth%
\newdimen\thicksize%
\newdimen\thinsize%
\newdimen\tablewidth%
\newif\ifcentertables%
\newif\ifendsize%
\newif\iffirstrow%
\newif\iftableinfo%
\newtoks\dbt%
\newtoks\hdtks%
\newtoks\savetks%
\newtoks\tableLETtokens%
\newtoks\tabletokens%
\newtoks\widthspec%
%
%
%
%
\tableinfotrue%
\catcode`\@=11
%
%
\def\tstrut{\vrule height3.1ex depth1.2ex width0pt}%
\def\and{\char`\&}
\def\tablerule{\noalign{\hrule height\thinsize depth0pt}}%
\thicksize=1.5pt
\thinsize=0.6pt
\def\thickrule{\noalign{\hrule height\thicksize depth0pt}}%
\def\ctr#1{\hfil\ #1\hfil}%
%
%
%
%
\tablewidth=-\maxdimen%
\spreadwidth=-\maxdimen%
\def\tabskipglue{0pt plus 1fil minus 1fil}%
%
%
\centertablestrue%
%
%
%
%
\parasize=4in%
\gdef\ARGS{########}
\gdef\headerARGS{####}
\def\@mpersand{&}
{\catcode`\|=13
\gdef\letbarzero{\let|0}
\gdef\letbartab{\def|{&&}}%
\gdef\letvbbar{\let\vb|}%
}
{\catcode`\&=4
\def\ampskip{&\omit\hfil&}
\catcode`\&=13
\let&0
\xdef\letampskip{\def&{\ampskip}}%
\gdef\letnovbamp{\let\novb&\let\tab&}
}
\def\begintable{
   \begingroup%
   \catcode`\|=13\letbartab\letvbbar%
   \catcode`\&=13\letampskip\letnovbamp%
   \def\multispan##1{
      \omit \mscount##1%
      \multiply\mscount\tw@\advance\mscount\m@ne%
      \loop\ifnum\mscount>\@ne \sp@n\repeat%
   }
   \def\|{%
      &\omit\widevline&%
   }%
   \ruledtable
}
\long\def\ruledtable#1\endtable{%
%
%
%
   \offinterlineskip
   \tabskip 0pt
   \def\widevline{\vrule width\thicksize}
   \def\endrow{\@mpersand\omit\hfil\crnorm\@mpersand}%
   \def\crthick{\@mpersand\crnorm\thickrule\@mpersand}%
   \def\crthickneg##1{\@mpersand\crnorm\thickrule
          \noalign{{\skip0=##1\vskip-\skip0}}\@mpersand}%
   \def\crnorule{\@mpersand\crnorm\@mpersand}%
   \def\crnoruleneg##1{\@mpersand\crnorm
          \noalign{{\skip0=##1\vskip-\skip0}}\@mpersand}%
   \let\nr=\crnorule
   \def\endtable{\@mpersand\crnorm\thickrule}%
   \let\crnorm=\cr
%
%
   \edef\cr{\@mpersand\crnorm\tablerule\@mpersand}%
   \def\crneg##1{\@mpersand\crnorm\tablerule
          \noalign{{\skip0=##1\vskip-\skip0}}\@mpersand}%
   \let\ctneg=\crthickneg
   \let\nrneg=\crnoruleneg
   \the\tableLETtokens
%
%
   \tabletokens={&#1}
%
%
   \countROWS\tabletokens\into\nrows%
   \countCOLS\tabletokens\into\ncols%
%
%
   \advance\ncols by -1%
   \divide\ncols by 2%
   \advance\nrows by 1%
%
%
   \iftableinfo %
      \immediate\write16{[Nrows=\the\nrows, Ncols=\the\ncols]}%
   \fi%
%
%
   \ifcentertables
      \ifhmode \par\fi
      \line{
      \hss
   \else %
      \hbox{%
   \fi
      \vbox{%
         \makePREAMBLE{\the\ncols}
         \edef\next{\preamble}
         \let\preamble=\next
         \makeTABLE{\preamble}{\tabletokens}
      }
      \ifcentertables \hss}\else }\fi
   \endgroup
   \tablewidth=-\maxdimen
   \spreadwidth=-\maxdimen
}
\def\makeTABLE#1#2{
   {
   \let\ifmath0
   \let\header0
   \let\multispan0
%
%
   \ncase=0%
   \ifdim\tablewidth>-\maxdimen \ncase=1\fi%
   \ifdim\spreadwidth>-\maxdimen \ncase=2\fi%
   \relax
%
   \ifcase\ncase %
      \widthspec={}%
   \or %
      \widthspec=\expandafter{\expandafter t\expandafter o%
                 \the\tablewidth}%
   \else %
      \widthspec=\expandafter{\expandafter s\expandafter p\expandafter r%
                 \expandafter e\expandafter a\expandafter d%
                 \the\spreadwidth}%
   \fi %
   \xdef\next{
      \halign\the\widthspec{%
      #1
      \noalign{\hrule height\thicksize depth0pt}
      \the#2\endtable
%
      }
   }
   }
   \next
}
\def\makePREAMBLE#1{
   \ncols=#1
   \begingroup
   \let\ARGS=0
   \edef\xtp{\widevline\ARGS\tabskip\tabskipglue%
   &\ctr{\ARGS}\tstrut}
   \advance\ncols by -1
   \loop
      \ifnum\ncols>0 %
      \advance\ncols by -1%
      \edef\xtp{\xtp&\vrule width\thinsize\ARGS&\ctr{\ARGS}}%
   \repeat
   \xdef\preamble{\xtp&\widevline\ARGS\tabskip0pt%
   \crnorm}
   \endgroup
}
\def\countROWS#1\into#2{
   \let\countREGISTER=#2%
   \countREGISTER=0%
   \expandafter\ROWcount\the#1\endcount%
}%
\def\ROWcount{%
   \afterassignment\subROWcount\let\next= %
}%
\def\subROWcount{%
   \ifx\next\endcount %
      \let\next=\relax%
   \else%
      \ncase=0%
      \ifx\next\cr %
         \global\advance\countREGISTER by 1%
         \ncase=0%
      \fi%
      \ifx\next\endrow %
         \global\advance\countREGISTER by 1%
         \ncase=0%
      \fi%
      \ifx\next\crthick %
         \global\advance\countREGISTER by 1%
         \ncase=0%
      \fi%
      \ifx\next\crnorule %
         \global\advance\countREGISTER by 1%
         \ncase=0%
      \fi%
      \ifx\next\crthickneg %
         \global\advance\countREGISTER by 1%
         \ncase=0%
      \fi%
      \ifx\next\crnoruleneg %
         \global\advance\countREGISTER by 1%
         \ncase=0%
      \fi%
      \ifx\next\crneg %
         \global\advance\countREGISTER by 1%
         \ncase=0%
      \fi%
      \ifx\next\header %
         \ncase=1%
      \fi%
      \relax%
      \ifcase\ncase %
         \let\next\ROWcount%
      \or %
         \let\next\argROWskip%
      \else %
      \fi%
   \fi%
   \next%
}
\def\counthdROWS#1\into#2{%
\dvr{10}%
   \let\countREGISTER=#2%
   \countREGISTER=0%
\dvr{11}%
\dvr{13}%
   \expandafter\hdROWcount\the#1\endcount%
\dvr{12}%
}%
\def\hdROWcount{%
   \afterassignment\subhdROWcount\let\next= %
}%
\def\subhdROWcount{%
   \ifx\next\endcount %
      \let\next=\relax%
   \else%
      \ncase=0%
      \ifx\next\cr %
         \global\advance\countREGISTER by 1%
         \ncase=0%
      \fi%
      \ifx\next\endrow %
         \global\advance\countREGISTER by 1%
         \ncase=0%
      \fi%
      \ifx\next\crthick %
         \global\advance\countREGISTER by 1%
         \ncase=0%
      \fi%
      \ifx\next\crnorule %
         \global\advance\countREGISTER by 1%
         \ncase=0%
      \fi%
      \ifx\next\header %
         \ncase=1%
      \fi%
\relax%
      \ifcase\ncase %
         \let\next\hdROWcount%
      \or%
         \let\next\arghdROWskip%
      \else %
      \fi%
   \fi%
   \next%
}%
{\catcode`\|=13\letbartab
\gdef\countCOLS#1\into#2{%
   \let\countREGISTER=#2%
   \global\countREGISTER=0%
   \global\multispancount=0%
   \global\firstrowtrue
   \expandafter\COLcount\the#1\endcount%
   \global\advance\countREGISTER by 3%
   \global\advance\countREGISTER by -\multispancount
}%
\gdef\COLcount{%
   \afterassignment\subCOLcount\let\next= %
}%
{\catcode`\&=13%
\gdef\subCOLcount{%
   \ifx\next\endcount %
      \let\next=\relax%
   \else%
      \ncase=0%
      \iffirstrow
         \ifx\next& %
            \global\advance\countREGISTER by 2%
            \ncase=0%
         \fi%
         \ifx\next\span %
            \global\advance\countREGISTER by 1%
            \ncase=0%
         \fi%
         \ifx\next| %
            \global\advance\countREGISTER by 2%
            \ncase=0%
         \fi
         \ifx\next\|
            \global\advance\countREGISTER by 2%
            \ncase=0%
         \fi
         \ifx\next\multispan
            \ncase=1%
            \global\advance\multispancount by 1%
         \fi
         \ifx\next\header
            \ncase=2%
         \fi
         \ifx\next\cr       \global\firstrowfalse \fi
         \ifx\next\endrow   \global\firstrowfalse \fi
         \ifx\next\crthick  \global\firstrowfalse \fi
         \ifx\next\crnorule \global\firstrowfalse \fi
         \ifx\next\crnoruleneg \global\firstrowfalse \fi
         \ifx\next\crthickneg  \global\firstrowfalse \fi
         \ifx\next\crneg       \global\firstrowfalse \fi
      \fi
\relax
      \ifcase\ncase %
         \let\next\COLcount%
      \or %
         \let\next\spancount%
      \or %
         \let\next\argCOLskip%
      \else %
      \fi %
   \fi%
   \next%
}%
\gdef\argROWskip#1{%
   \let\next\ROWcount \next%
}
\gdef\arghdROWskip#1{%
   \let\next\ROWcount \next%
}
\gdef\argCOLskip#1{%
   \let\next\COLcount \next%
}
}
}
\def\spancount#1{
   \nspan=#1\multiply\nspan by 2\advance\nspan by -1%
   \global\advance \countREGISTER by \nspan
   \let\next\COLcount \next}%
\def\dvr#1{\relax}%
\def\header#1{%
\dvr{1}{\let\cr=\@mpersand%
\hdtks={#1}%
\counthdROWS\hdtks\into\hdrows%
\advance\hdrows by 1%
\ifnum\hdrows=0 \hdrows=1 \fi%
\dvr{5}\makehdPREAMBLE{\the\hdrows}%
\dvr{6}\getHDdimen{#1}%
{\parindent=0pt\hsize=\hdsize{\let\ifmath0%
\xdef\next{\valign{\headerpreamble #1\crnorm}}}\dvr{7}\next\dvr{8}%
}%
}\dvr{2}}
\def\makehdPREAMBLE#1{
\dvr{3}%
\hdrows=#1
{
\let\headerARGS=0%
\let\cr=\crnorm%
\edef\xtp{\vfil\hfil\hbox{\headerARGS}\hfil\vfil}%
\advance\hdrows by -1
\loop
\ifnum\hdrows>0%
\advance\hdrows by -1%
\edef\xtp{\xtp&\vfil\hfil\hbox{\headerARGS}\hfil\vfil}%
\repeat%
\xdef\headerpreamble{\xtp\crcr}%
}
\dvr{4}}
\def\getHDdimen#1{%
\hdsize=0pt%
\getsize#1\cr\end\cr%
}
\def\getsize#1\cr{%
\endsizefalse\savetks={#1}%
\expandafter\lookend\the\savetks\cr%
\relax \ifendsize \let\next\relax \else%
\setbox\hdbox=\hbox{#1}\newhdsize=1.0\wd\hdbox%
\ifdim\newhdsize>\hdsize \hdsize=\newhdsize \fi%
\let\next\getsize \fi%
\next%
}%
\def\lookend{\afterassignment\sublookend\let\looknext= }%
\def\sublookend{\relax%
\ifx\looknext\cr %
\let\looknext\relax \else %
   \relax
   \ifx\looknext\end \global\endsizetrue \fi%
   \let\looknext=\lookend%
    \fi \looknext%
}%
%
%
\def\tablelet#1{%
   \tableLETtokens=\expandafter{\the\tableLETtokens #1}%
}%
\catcode`\@=12
%
\def\NSVZ{{\rm NSVZ}}

\def\DREDp{{\rm DRED}'}
\def\sy{supersymmetry}
\def\sic{supersymmetric}
\def \inparg{\leftskip = 40 pt\rightskip = 40pt}
\def \outparg{\leftskip = 0 pt\rightskip = 0pt}
\def\vev#1{\mathopen\langle #1\mathclose\rangle }
\thicksize=0.7pt
\thinsize=0.5pt
\def\ctr#1{\hfil $\,\,\,#1\,\,\,$ \hfil}  
\def\tstrut{\vrule height 2.7ex depth 1.0ex width 0pt}

\def\pa{\partial}
\def\semi{;\hfil\break}
\def\MSSMnu{\hbox{MSSM}^{\nu}}
\def\frak#1#2{{\textstyle{{#1}\over{#2}}}}
\def\frakk#1#2{{{#1}\over{#2}}}

\def\hbar{{\overline h}{}}
\def\mbar{{\overline{m}}}
\def\nbar{{\overline{n}}}
\def\qbar{{\overline{q}}}

\def\mbar{{\overline m}{}}

\def\semi{;\hfil\break}
\def\npb{{Nucl.\ Phys.\ }{\bf B}}
\def\prd{{Phys.\ Rev.\ }{\bf D}}
\def\prl{Phys.\ Rev.\ Lett.\ }
\def\plb{{Phys.\ Lett.\ }{\bf B}}

\def\Ccal{{\cal C}}

\def\Rcal{{\cal R}}
\def\Ycal{{\cal Y}}
\def\Ocal{{\cal O}}

\def\Wtil{\tilde W}
  
\def\Ytil{\tilde Y}
\def\tautil{\tilde\tau}
\def\chitil{\tilde\chi}
\def\nutil{\tilde\nu}
\def\mutil{\tilde\mu}

\def\btil{\tilde b}
\def\ctil{\tilde c}
\def\dtil{\tilde d}
\def\etil{\tilde e}
\def\gtil{\tilde g}
\def\ltil{\tilde l}

\def\qtil{\tilde q}
\def\stil{\tilde s}
\def\ttil{\tilde t}
\def\util{\tilde u}

\def\TeV{{\rm TeV}}
\def\GeV{{\rm GeV}}
\def\eV{{\rm eV}}
{\nopagenumbers
\line{\hfil LTH 568}
\line{\hfil hep-ph/0301163}
\line{\hfil Revised Version}
\vskip .5in
\centerline{\titlefont Yukawa Textures and}
\centerline{\titlefont Anomaly Mediated Supersymmetry Breaking}
\vskip 1in
\centerline{\bf I.~Jack and  D.R.T.~Jones}
\bigskip
\centerline{\it Dept. of Mathematical Sciences,
University of Liverpool, Liverpool L69 3BX, U.K.}
\vskip .3in

We present a detailed analysis of how  a mixed-anomaly-free  $U_1$
symmetry can be used to both  resolve the slepton mass problem
associated with  Anomaly Mediated Supersymmetry Breaking and generate
the  fermion mass  hierarchy via the Froggatt-Nielsen  mechanism.  
Flavour changing neutral currents problems are evaded by a  specific
form of the Yukawa textures. 

\Date{Jan  2003}}

\newsec{Introduction}

An explanation for the existence of the three generations of quarks and 
leptons with their widely dispersed masses remains one of the most
significant  problems in fundamental physics. One plausible  way  they
may emerge from a more fundamental theory is via Yukawa textures
associated  with a $U_1$ symmetry, either global or gauged at some
higher scale
\ref\cdfr{C.D.~Froggatt and H.B.~Nielsen, \npb 147 (1979) 277
\semi 
M.~Leurer, Y.~Nir and N.~Seiberg,  \npb 398 (1993) 319
\semi 
P.~Ramond, R.G.~Roberts and  G.G.~Ross,
\npb 406 (1993) 19}. This 
idea has been much studied both with anomaly-free and anomalous $U_1$'s. 
In this paper we investigate  
Yukawa textures in the context of a 
specific framework for the origin of soft supersymmetry breaking 
within the MSSM,
known as Anomaly Mediated Supersymmetry Breaking (AMSB)
\ref\lisa{L.~Randall and R.~Sundrum, \npb 557 (1999) 79}%
\nref\glmr{G.F.~Giudice, M.A.~Luty, H.~Murayama and  R.~Rattazzi,
JHEP 9812 (1998) 27}%
--\ref\jjrg{I.~Jack and D.R.T.~Jones, \plb 465 (1999) 148}.
Direct application of the AMSB solution  to the MSSM leads, unfortunately,  to
negative $(\hbox{mass})^2$ sleptons. 
A number of possible solutions to this problem have been discussed;  here we
concentrate on proposals 
\ref\jjfi{I.~Jack and D.R.T.~Jones, \plb 482 (2000) 167}
\ref\jjrp{I.~Jack and D.R.T.~Jones, \plb 491 (2000) 151} which 
require the existence of an additional  $U_1$ 
symmetry; in the first case (Case~(FI)) a ``normal'' $U'_1$ 
(commuting with \sy), and in the second 
case (Case~($\cal{R}$)) a $U'_1$ associated with an ${\cal R}$-symmetry
\ref\appp{A.~Pomarol and  R.~Rattazzi, JHEP 9905 (1999) 013\semi
M.A.~Luty and R.~Rattazzi, JHEP 9911 (1999) 001}.
Both cases permit additional contributions to the scalar 
masses which preserve  the exact RG
invariance of the AMSB solution, providing the $U'_1$ has no 
linear mixed  anomalies with any gauge group of the theory. 
In the MSSM context this amounts to the requirement that the 
$U'_1 (SU_3)^2$, $U'_1 (SU_2)^2$ and $U'_1 (U_1)^2$ anomalies cancel.
The $U'_1$ need not in
fact be gauged, though of course the vanishing of these anomalies 
suggests that it may be. We will therefore also impose 
cancellation of $U_1 (U'_1)^2$ anomalies, so that a MSSM singlet 
sector would suffice to render $U'_1$ anomaly free. It is very natural 
to use the same  $U'_1$ symmetry to both solve the slepton mass problem 
{\it and\/} generate the Yukawa textures.

With regard to Case~(FI), the MSSM
in fact  admits two generation-independent, 
mixed-anomaly-free  $U_1$ groups, the existing $U_1^Y$ and another
(which could be chosen to be  $U_1^{B-L}$
\ref\nahk{N.~Arkani-Hamed et al, JHEP 0102 (2001) 041}\ 
or a linear
combination of it and $U_1^Y$\jjfi). The existence 
of these two independent $U_1$'s indeed enables us to resolve the slepton 
problem and predict a distinctive sparticle spectrum with 
characteristic mass sum rules, as described in Ref.~\jjfi.
This scenario has the advantage of incorporating natural suppression
of flavour-changing effects in both hadronic and leptonic sectors.  
It provides no insight into the flavour problem, however, and does not 
accommodate neutrino masses  in an elegant way. 

The MSSM does not admit a 
generation-independent ${\Rcal}$ symmetry (although one can be arranged 
using additional matter fields, see 
Ref.~\ref\KitazawaFT{
N.~Kitazawa, N.~Maru and N.~Okada,
\prd 63  (2001) 015005; \npb 586 (2000) 261;  \prd 62  (2000) 077701}), and 
so there is no analogous treatment to Case~(FI). 
In Ref.~\jjrp\ we argued that in conjunction with the Froggatt-Nielsen
(FN) mechanism\cdfr, use of a generation-dependent ${\Rcal}$-symmetry 
could combine the desirable features of the AMSB scenario with 
an explanation for the flavour hierarchy. The form of Yukawa textures 
adopted in Ref.~\jjrp\ was motivated by the limits imposed by 
flavour changing neutral currents (FCNC); it required, however, some 
fine tuning to reproduce the flavour hierarchy and the CKM matrix. 
In Ref.~\ref\jjwb{I.~Jack, D.R.T.~Jones and R.~Wild, \plb 535 (2002) 193}, 
motivated by the now overwhelming evidence for massive neutrinos
\foot{Direct extension of Case~(FI) to include massive neutrinos
is possible  using Dirac mass terms, but this is very unattractive 
as it provides no explanation for the extreme lightness of the neutrinos. 
The seesaw mechanism provides just such an explanation, and most elegantly; 
but involves Majorana masses for right-handed
neutrinos, which evidently break $U_1^{B-L}$. 
For some alternative ideas about massive neutrinos in this context 
see Ref.~\nahk.}, we 
considered a generation-dependent $U_1$ in the Case~(FI) context, and 
found a different texture which both naturally reproduces the flavour hierarchy 
and leads to an acceptable CKM matrix. Our purpose here is to describe this 
scenario in more detail, explore alternatives, and extend the discussion 
to encompass Case~$({\cal R})$. We show how the existence of a mass sum 
involving the Higgs bosons constrains our $U'_1$ charge assignments, and 
exhibit sparticle mass spectra for the various possible scenarios. 
In section~2 we review AMSB, in section~3 we describe briefly the MSSM 
generalised to incorporate massive neutrinos via the see-saw mechanism
(which we term the $\MSSMnu$); in sections~4-7 we analyse the constraints 
imposed by anomaly cancellation  and in section~8 
we pursue the experimental consequences of our preferred choice of textures.

\newsec{Anomaly mediation}

Consider a \sic\ theory with superpotential
\eqn\newW{
W(\phi) = \frak{1}{2}{\mu}^{ij}\phi_i\phi_j + \frak{1}{6}Y^{ijk}
\phi_i\phi_j\phi_k,}
and soft \sy-breaking terms as follows:
\eqn\newV{V_{\hbox{soft}} = 
\left( \frak{1}{2}b^{ij}\phi_i\phi_j 
+ \frak{1}{6}h^{ijk}\phi_i\phi_j\phi_k  + \half M \lambda\lambda
+\hbox{c.c.}\right)
+(m^2)^i{}_j\phi_i\phi^j. }
The anomaly mediation approach to the MSSM begins with 
the following relations:
\eqna\amsbsome$$\eqalignno{
M &= M_0{\beta_g\over g},&\amsbsome a\cr
h^{ijk} &= -M_0\beta_Y^{ijk}, &\amsbsome b\cr
(m^2)^i{}_j &= \frak{1}{2}|M_0|^2\mu{d\gamma^i{}_j\over{d\mu}}, 
&\amsbsome c\cr}$$
which are RG invariant to all orders of perturbation theory.
 (In appendix~A we provide a summary of the most general set 
of such relations, for a theory including gauge singlets.) 

Eq.~\amsbsome{c}\ leads to tachyonic sleptons;
most studies have dealt with the so-called mAMSB, produced by replacing 
it (at the unification scale) with 
\eqn\marrrgnota{m^2 = m^2_{\hbox{AMSB}} + m_0^2,}
that is 
\eqn\massrgnot{
(m^2)^i{}_j = \frak{1}{2}|M_0|^2\mu{d\gamma^i{}_j\over{d\mu}}
+m_0^2\delta^i{}_j,} where $m_0^2$ is constant. 
This procedure, however, destroys the RG invariance (and hence the UV 
insensitivity) of the relation. Much more elegant, in our opinion, are 
the following two possibilities:

\subsec{Case~(FI): The Fayet-Iliopoulos solution}

Here we replace Eq.~\amsbsome{c}\ with:
\eqn\massrgb{
(m^2)^i{}_j = \frak{1}{2}|M_0|^2\mu{d\gamma^i{}_j\over{d\mu}}
+\sum_a \zeta_a (\Ycal_a)^i{}_j,} where $\zeta_a$, $(\Ycal_a)^i{}_j$ are 
constants, satisfying the following relations:

\eqna\amsbsomea$$\eqalignno{
(\Ycal_a)^i{}_lY^{ljk}+(\Ycal_a)^j{}_lY^{ilk}+(\Ycal_a)^k{}_lY^{ijl} & =  0
 & \amsbsomea a\cr
\tr [\Ycal_a C(R)] & = 0.
 & \amsbsomea b\cr}$$
Here $C(R)$ is the quadratic gauge Casimir for the chiral multiplet. 
These constraints follow from demanding that $m^2$ be RG invariant,
but clearly correspond to requiring that each $\Ycal$ 
correspond to an abelian symmetry of the superpotential (Eq.~\amsbsomea{a}), 
such that  all anomalies linear in  
$\Ycal$ and quadratic in {\it gauged\/} symmetries vanish (Eq.~\amsbsomea{b}). 
It is interesting that the 
latter requirement derives from the $X$-function in the $\beta$-function for 
$m^2$; this  function, whose existence was first remarked in 
Ref.~\ref\jjpa{I.~Jack, D.R.T.~Jones and A.~Pickering,
\plb426 (1998) 73}, 
was related recently to anomalies in 
Ref.~\ref\kraus{E.~Kraus and D.~Stockinger,
\prd 65 (2002)  105014}. 
The $\zeta_a$-terms in Eq.~\massrgb\ correspond 
precisely to the contributions to the scalar masses from 
Fayet-Iliopoulos (FI) $D$-terms, after elimination of the 
auxiliary $D$-fields  using 
their equations of motion.    
Note that the mixed anomaly cancellation requirement 
rules out anomaly cancellation via the Green-Schwarz 
mechanism \ref\grsch{M.~Green and J.~Schwarz, \plb 149 (1984) 117}. 
The simplest realisation of the FI scenario  is to have two $\zeta$'s, 
$\zeta$, $\zeta'$, the first corresponding to the standard model $U_1$ and 
the second to a $U'_1$ the possible form of which 
we will discuss in detail later.

\subsec{Case~$({\cal R})$: The ${\Rcal}$-symmetry solution}

Here we have  

\eqn\Rsol{
(m^2)^i{}_j = \frak{1}{2}|M_0|^2\mu{d\gamma^i{}_j\over{d\mu}} + 
m_0^2 (\gamma^i{}_j + \qbar_i \delta^i{}_j)}
where $m_0^2$ and $\qbar_i$ are constants, as long as
a set $\qbar_i$ exists that satisfy the following constraints:
\eqna\qrels$$\eqalignno{
(\qbar_i+\qbar_j+\qbar_k)Y^{ijk} &= 0 &\qrels a\cr
2\Tr\left[\qbar C(R)\right] + Q &= 0, & \qrels b\cr}$$
where $Q$ is the one loop $\beta_g$ coefficient. 
It is easy to show\appp\
that Eq.~\qrels{}\ corresponds precisely to requiring
that the theory have a non-anomalous
${\Rcal}$-symmetry (which we denote ${\cal R}$, to
avoid confusion with our notation $R$ for group representations),
where if we set
\eqn\qcharg{\qbar_i = 1 -\frak{3}{2}r_i,}
then Eq.~\qrels{a}\ corresponds to $(r_i+r_j+r_k)Y^{ijk} = 2Y^{ijk}$,
which is the conventional ${\cal R}$-charge normalisation. In the 
rest of the paper we will work with the fermionic $\Rcal$ charges, 
$q_i = r_i - 1$. 
The relation Eq.~\Rsol\ generalises easily to the case of several 
$\Rcal$ symmetries but the simplest possibility, which suffices 
to deal with the slepton mass problem, is to have one only, which 
we will call $U_1^{\Rcal}$.

\newsec{The superpotential and neutrino masses}

The $\MSSMnu$ is defined by the superpotential 
\eqn\supernu{\eqalign{W &= H_2 Q Y_u u^c  + H_1 Q Y_d d^c + H_1 L Y_e e^c
+ H_2 L Y_{\nu}\nu^c\cr 
&+\mu H_1 H_2 + \frak{1}{2}(\nu^c)^T M_{\nu^c}\nu^c,\cr}}
where $Y_u$, $Y_d$, $Y_e$ are $3\times 3$ Yukawa matrices, and 
$Y_{\nu}$ is $3 \times n_{\nu}$, where $n_{\nu}$ is the number of RH 
neutrinos.  
We will neglect CP-violation and assume that all parameters 
in Eq.~\supernu\ are real. The light neutrino mass matrix $m_{\nu}$ is
generated by the seesaw mechanism, 
\eqn\numasses{
m_{\nu} = m_D M_{\nu^c}^{-1} (m_D)^T}
where $m_D = v_2Y_{\nu}$ is the Dirac $\nu$-mass matrix.

The quark and charged lepton matrices are diagonalised as follows: 
\eqn\qldiag{
Y_u^{\hbox{diag}} = U_u^{T} Y_u V_u}
(similarly for $Y_d, Y_e$), so that the CKM matrix is given by 
\eqn\ckmdef{CKM = U_u^{T} U_d.}
The neutrino mixing matrix $U_{MNS}$ relating the mass eigenstate basis to the 
basis in which the leptonic charged currents are flavour-diagonal, 
i.e. 
\eqn\numix{\pmatrix{\nu_e\cr\nu_{\mu}\cr\nu_{\tau}\cr} = 
U_{MNS}\pmatrix{\nu_1\cr\nu_2\cr\nu_3\cr}}
is given by 
\eqn\numix{
U_{MNS} = U_{e}^{T} U_{\nu},}   
where 
\eqn\nudiag{
m_{\nu}^{\hbox{diag}} = U_{\nu}^{T} m_{\nu} U_{\nu}.}

Existing oscillation data suggests non-zero neutrino masses with 
large mixing. Super-Kamiokande results are consistent with 
$\nu_{\mu} \leftrightarrow \nu_{\tau}$ oscillations with 
$\delta m^2 \sim 10^{-3}\eV^2$ and maximal mixing;
while the large-angle  MSW solution to the solar oscillation data 
is consistent with $\nu_{\mu} \leftrightarrow \nu_{e}$ mixing with 
$\delta m^2 \sim 10^{-5}\eV^2$ and large (not quite maximal) mixing. 
Finally the CHOOZ reactor experiment and the Palo Verde experiment 
suggest that $U_{e3}$ is small. 

Thus we seek a $U_{MNS}$ such that  
\eqn\numix{
U_{MNS} = 
\pmatrix{c_{12}c_{31}&s_{12}c_{31}&s_{31}\cr
-s_{12}c_{23}-c_{12}s_{23}s_{31}&c_{12}c_{23}-s_{12}s_{23}s_{31}
&s_{23}c_{31}\cr   s_{12}s_{23}-c_{12}c_{23}s_{31}
&-c_{12}s_{23}-s_{12}c_{23}s_{31}&c_{23}c_{31}\cr}}
with $\sin^2 2\theta_{12}\sim 0.75$ and $\sin^2 2\theta_{23}\sim 1$ and 
$\sin^2 2\theta_{31}\sim 0$. 

Examples from the literature of favoured structures are:

\eqn\numixa{
U_{MNS} = 
\pmatrix{\cos\phi&-\sin\phi&0\cr
\sin\phi/\sqrt{2}
&\cos\phi/{\sqrt{2}}&-1/\sqrt{2}\cr\sin\phi/\sqrt{2}
&\cos\phi/\sqrt{2}&1/{\sqrt{2}}\cr}}
or\ref\HarrisonKP{
P.F.~Harrison and W.G.~Scott,
\plb 535 (2002) 163 
}
\eqn\numixb{
U_{MNS} =
\pmatrix{\sqrt{\frak{2}{3}}\cos\phi&\sqrt{\frak{1}{3}}&
\sqrt{\frak{2}{3}}\sin\phi\cr
-\frak{\cos\phi}{\sqrt{6}}-\frak{\sin\phi}{\sqrt{2}}&\sqrt{\frak{1}{3}}
&\frak{\cos\phi}{\sqrt{2}}-\frak{\sin\phi}{\sqrt{6}}\cr 
-\frak{\cos\phi}{\sqrt{6}}+\frak{\sin\phi}{\sqrt{2}}
&\sqrt{\frak{1}{3}}& -\frak{\cos\phi}{\sqrt{2}} -\frak{\sin\phi}{\sqrt{6}}\cr}}
\break

We will discuss later to what extent our framework predicts 
(or at least accommodates) results of this general nature.  

\newsec{The Yukawa Textures}

In this section we will discuss the form of $Y_{u,d,e}$, postponing 
$Y_{\nu}$ till later. 
We seek to reproduce the well-known hierarchies
\ref\rgrss{ P.~Ramond, R.G.~Roberts and G.G.~Ross, \npb 406 (1993) 19 }
\eqn\massrats{
m_{\tau}:m_{\mu}:m_e = m_b:m_s:m_d = 1:\lambda^2:\lambda^4,
\quad\hbox{and}\quad
m_t:m_c:m_u = 1:\lambda^4:\lambda^8}
(where $\lambda\approx 0.22$), 
an acceptable CKM matrix (without too much fine tuning), 
and neutrino masses and mixings consistent with current observations. 

The fundamental assumption we shall make is that 
most of the Yukawa interactions are generated 
via the Froggatt-Nielsen (FN)
mechanism\cdfr: specifically, from
higher dimension terms involving $\hbox{MSSM}^{\nu}$ singlet fields
$\theta_{u,d,e}$ with $U'_1$ or $U^{\Rcal}_1$ charges $-Q_u$, $-Q_d$, 
and $-Q_e$,  via terms such as
$H_2 Q_i u^c_j (\frakk{\theta_u}{M_{\theta}})^{a_{ij}}$, 
where $M_{\theta}$ represents
the scale of new physics; we will assume that $M_{\theta} \geq M_{\nu^c}$. 
We choose to normalise charges so that 
$Q_u = 1$. For our principal development 
we will assume that each Yukawa matrix $Y_{u,d,e}$ gains 
its texture from a  {\it particular\/} $\theta$-charge and that the vevs 
of the various  $\theta$-charges are approximately the same.
Assigments such that this scenario is (in a sense we will define) 
{\it natural\/}  will be discussed presently. 
\foot{If the $U'_1$ is in fact gauged then this implicitly assumes
that there exists a D-flat direction corresponding to the 
$M_{\theta}$ scale with all the vevs of the $\theta\hbox{s}$ involved 
in texture generation approximately the same.}
It follows 
at once from gauge invariance that the textures take the following form: 
\eqn\texts{\eqalign{
Y_u &\sim \phantom{\lambda^{\alpha_e}}
\pmatrix{\lambda^{p_u}&\lambda^{x_u}&\lambda^{y_u}\cr  
\lambda^{a_u}&\lambda^{q_u}&\lambda^{z_u}\cr
\lambda^{b_u}&\lambda^{c_u}&1}, 
\quad
Y_d \sim \lambda^{\alpha_d}
\pmatrix{\lambda^{p_d}&\lambda^{x_d}&\lambda^{y_d}\cr  
\lambda^{a_d}&\lambda^{q_d}&\lambda^{z_d}\cr\lambda^{b_d}&\lambda^{c_d}&1},\cr
 Y_e &\sim \lambda^{\alpha_e}
\pmatrix{\lambda^{p_e}&\lambda^{x_e}&\lambda^{y_e}\cr  
\lambda^{a_e}&\lambda^{q_e}&\lambda^{z_e}\cr\lambda^{b_e}
&\lambda^{c_e}&1}
\cr}} 
where (in both Case~(FI) and Case~(${\cal R}$))
\eqn\exprels{\eqalign{a_u &= p_u+q_u-x_u\cr
b_u &= p_u-y_u\cr
c_u &= x_u-y_u\cr
z_u &= q_u+y_u-x_u,\cr}}
with similar relations for $a_{d,e}$ etc. Given that $m_{\tau} \sim m_b$ 
and   $m_b/m_t \sim \lambda^3$ we might  expect $\alpha_d \sim \alpha_e$ 
and $\tan\beta \sim \lambda^{\alpha_d -3}$. Hence for $\tan\beta \sim 10$, 
for example, we would then expect $\alpha_d = 1$ or $\alpha_d =2$. 
It might also be, however, that the hierarchy $m_b / m_t$ has a different
origin.  

Gauge invariance of the Yukawas also provides 
relationships among the various charges. 
We will first give these relationships for the $U'_1$ charges in Case~(FI).
Denoting the $U'_1$ charges of the supermultiplets 
$Q_i, L_i, u^c_i, d^c_i, e^c_i, 
H_1, H_2$ as $q_i, L_i, u_i, d_i, e_i, h_1, h_2$, we 
have:  
\eqn\exprelsa{\eqalign{
q_1 &=p_u-u_1-h_2\cr
q_2 &=a_u-u_1-h_2\cr
q_3 &=b_u-u_1-h_2\cr
L_1 &= (p_e+\alpha_e)Q_e-e_1-h_1\cr
L_2 &=  (a_e+\alpha_e)Q_e-e_1-h_1\cr
L_3 &= (b_e+\alpha_e)Q_e-e_1-h_1,\cr
\cr}}
\eqn\exprelsb{\eqalign{
u_2 &=u_1+q_u-a_u   \cr
u_3 &=u_1-b_u\cr
d_1 &=(\alpha_d+p_d)Q_d - p_u + u_1-h_1+h_2\cr
d_2 &=(\alpha_d+ q_d)Q_d-a_u+u_1-h_1+h_2\cr
d_3 &= \alpha_dQ_d-b_u+u_1-h_1+h_2\cr
e_2 &= (q_e-a_e)Q_e+e_1\cr
e_3 &= -Q_e b_e+e_1\cr
\cr}}
and also the following relations:
\eqn\exprelsc{\eqalign{
a_d &= (a_u-p_u+Q_d p_d)/Q_d\cr
b_d &=   (b_u+Q_d p_d-p_u)/Q_d\cr
c_d &=   (b_u+Q_d q_d-a_u)/Q_d\cr
x_d &= (p_u+Q_d q_d-a_u)/Q_d\cr
y_d &= (p_u-b_u)/Q_d\cr
z_d &= (a_u-b_u)/Q_d.\cr
\cr}}

Case~(${\cal R})$ differs because the superpotential has 
non-zero $\Rcal$-charge which as usual we take to be 2. 
Then, in terms of the $U^{\Rcal}_1$ {\it fermionic\/} charges,
we have (instead of Eq.~\exprelsa):
\eqn\exprelsr{\eqalign{
q_1 &=p_u-u_1-h_2 - 1 \cr
q_2 &=a_u-u_1-h_2 - 1 \cr
q_3 &=b_u-u_1-h_2 - 1 \cr
L_1 &= (p_e+\alpha_e)Q_e-e_1-h_1 - 1 \cr
L_2 &=  (a_e+\alpha_e)Q_e-e_1-h_1 - 1 \cr
L_3 &= (b_e+\alpha_e)Q_e-e_1-h_1- 1, \cr
}}
while Eqs.~\exprelsb, \exprelsc\ are unaffected.

Cancellation of mixed anomalies for 
$(SU_3)^2 U'_1$, $(SU_2)^2 U'_1$ and $(U_1)^2 U'_1$ leads to the conditions
(in Case~(FI))
\eqna\anomcanc$$\eqalignno{
A_3 &= \sum_{i=1}^3  (2q_i+u_i+d_i) = 0 & \anomcanc a\cr
A_2 &= \Delta+ \sum_{i=1}^3 (L_i+3q_i) = 0 & \anomcanc b\cr
A_1 &= 3\Delta + \sum_{i=1}^3 (3L_i +q_i +8u_i+2d_i + 6e_i)  = 0 
& \anomcanc c\cr}$$
where we have set $h_1 = \Delta - h_2$, or in Case~(${\cal R})$:

\eqna\anomcanr$$\eqalignno{
A_3 &= 6 + \sum_{i=1}^3  (2q_i+u_i+d_i) = 0 & \anomcanr a\cr
A_2 &= 4 + \Delta+ \sum_{i=1}^3 (L_i+3q_i) = 0 & \anomcanr b\cr
A_1 &= 3\Delta + \sum_{i=1}^3 (3L_i +q_i +8u_i+2d_i + 6e_i)  = 0, 
& \anomcanr c\cr}$$
where the additional contributions in Eq.~\anomcanr{a,b}\ are 
due to gauginos. 

Cancellation of  $ U_1 (U'_1)^2$ or $ U_1 (U^{\Rcal}_1)^2$  anomalies 
leads to the further condition (valid in both cases)
\eqn\anomquad{
A_Q = -h_1^2+h_2^2+\sum_i^3 (e_i^2-L_i^2+q_i^2-2u_i^2+d_i^2).
}
MSSM singlet fields (such as $\nu^c_i$ and the $\theta$-fields) 
do not contribute to Eqs.~\anomcanc{}-\anomquad. They {\it do\/} contribute 
to $U'_1$-gravitational and $(U'_1)^3$ anomalies, which are proportional to 
the following expressions  
(which we include for completeness but the vanishing of which we 
do not impose). In the FI case: 
\eqn\anomcubic{A_C = 
 2(h_1^3+h_2^3)+\sum_{i=1}^3 (e_i^3+2L_i^3+6q_i^3+3u_i^3+3d_i^3) +\sum s_j^3,}
and 
\eqn\anomgrav{A_G = 
 2\Delta+\sum_{i=1}^3 (e_i+2L_i+6q_i+3u_i+3d_i) + \sum s_j,}
and in the $\Rcal$ case:
\eqn\anomcubicR{A_C = 
 2(h_1^3+h_2^3)+16+\sum_{i=1}^3 (e_i^3+2L_i^3+6q_i^3+3u_i^3+3d_i^3)+\sum s_j^3,}
and 
\eqn\anomgravR{A_G = 
 2\Delta -8+\sum_{i=1}^3 (e_i+2L_i+6q_i+3u_i+3d_i)+ \sum s_j,}
where we have assumed a singlet sector with charges $s_i$ which would include 
$\nu^c_i$ and the $\theta$-fields. Note the gravitino contributions 
to $A_{C,G}$ in the $\Rcal$ case\ref\ChamseddineGB{
A.H.~Chamseddine and H.K.~Dreiner,
\npb 458  (1996) 65\semi
D.J.~Castano, D.Z.~Freedman and C.~Manuel, \npb 461 (1996) 50}.

\newsec{The Wolfenstein textures}

Here we explore whether we can obtain the Wolfenstein texture 
for the CKM  matrix,
\eqn\ckmwolf{CKM_W  \sim \pmatrix{1&\lambda&\lambda^3\cr
\lambda&1&\lambda^2\cr \lambda^3&\lambda^2&1}}
in the light of the constraints imposed in the previous section. 
There is a considerable literature on this subject, both 
with anomaly-free and anomalous $U'_1$ symmetries; for a recent 
example see Ref.~\ref\TanakaPY{
S.~Tanaka,
\plb 480 (2000)  296
}. 

A matrix of the form e.g. $Y_u$ in Eq.~\texts\ has one eigenvalue $O(1)$
and two of orders $O(\lambda^{\min\{p_u,q_u,x_u,a_u\}})$, 
$O(\lambda^{\max\{p_u,q_u,x_u,a_u\}})$ respectively. Moreover, 
the Wolfenstein texture for the CKM  matrix is
obtained if the right-hand columns of $Y_u$ and $Y_d$ are both of the
form 
$\pmatrix{\lambda^3 & \lambda^2 & 1\cr}^T$, 
and if $x_{u,d}\ge3$. 
If we require mass 
hierarchies of the form Eq.~\massrats, then the only possible textures 
satisfying these conditions (together with Eqs.~\exprels) are of the form    
\eqn\textstan{
Y_u \sim \pmatrix{\lambda^{8}&\lambda^{5}&\lambda^{3}\cr  
\lambda^{7}&\lambda^{4}&\lambda^{2}\cr
\lambda^{5}&\lambda^{2}&1}, 
\quad
Y_d \sim \lambda^{\alpha_d}
\pmatrix{\lambda^{4}&\lambda^{3}&\lambda^{3}\cr  
\lambda^{3}&\lambda^{2}&\lambda^{2}\cr\lambda&1&1}.}
All the relevant conditions from Eq.~\exprelsc\
are then satisfied by taking $Q_d=1$. 

We can then solve the linear anomaly constraints Eq.~\anomcanc{}\ 
for $\Delta$, $e_1$ and $Q_e$, yielding (in case~(FI)):
\eqna\qecons$$\eqalignno{
\Delta &= \alpha_d + 6   & \qecons a\cr 
Q_e &= 2 \alpha_d/(3 \alpha_e+p_e+q_e) & \qecons b\cr
u_1 &= -2\alpha_d/9 +16/3-2 h_2/3-e_1/3
+Q_e (p_e + a_e +b_e + 3\alpha_e )/9
 & \qecons c\cr}$$
while in case~(${\cal R})$ we find from Eq.~\anomcanr{}\ that
\eqna\qeconsR$$\eqalignno{
\Delta &= \alpha_d + 6   & \qeconsR a\cr 
Q_e &= 2 ( 3 +\alpha_d)/(3 \alpha_e+p_e+q_e) & \qeconsR b\cr
u_1 &= 40/9-2\alpha_d/9-2h_2/3-e_1/3+Q_e (p_e + a_e +b_e + 3\alpha_e )/9
& \qeconsR c\cr}$$
In both cases we necessarily have a texture-generated 
$\mu$-term, related to $M_{\theta}$ by 
$\mu \sim M_{\theta} (\lambda)^{\alpha_d +6}$, 
assuming that the $\theta$ responsible 
for it has the same $U'_1$ charge as $\theta_{u,d}$.

Imposing Eq.~\qecons{}\ or Eq.~\qeconsR{}\ renders Eq.~\anomquad\ 
linear in $h_2$,   
so we solve Eq.~\anomquad\ for $h_2$, obtaining in Case~(FI):
\eqn\aqsoln{\eqalign{
h_2 &=  
(116 - 2 Q_e p_e \alpha_d - 2 Q_e a_e \alpha_d + 32 \alpha_d + 12 e_1
 - 4 Q_e p_e \cr& - 12 Q_e \alpha_e - 4 Q_e a_e - 4 Q_e b_e + Q_e^2 p_e^2
 + 3 Q_e^2\alpha_e^2  + 2 Q_e^2 p_e \alpha_e \cr& 
- 2 Q_e p_e e_1 - 6 Q_e \alpha_e e_1
 + 2 Q_e^2 a_e \alpha_e + 2 Q_e^2  b_e \alpha_e + 2 Q_e^2  q_e a_e\cr&
 - 2 Q_e q_e e_1 - Q_e^2 q_e^2 + 4 \alpha_d^2 - 6 Q_e \alpha_e \alpha_d
 - 2 Q_e b_e \alpha_d + 6 e_1 \alpha_d)/(4(6 + \alpha_d))\cr}}
and in Case~(${\cal R})$: 
\eqn\aqsolnr{\eqalign{
h_2 &=    
(-6 Q_e a_e \alpha_d - 6 Q_e b_e \alpha_d - 18 Q_e \alpha_e \alpha_d
+ 6 Q_e^2  p_e \alpha_e \cr& + 118 \alpha_d - 6 Q_e p_e \alpha_d 
- 6 Q_e p_e e_1 - 20 Q_e a_e - 60 Q_e \alpha_e 
\cr& - 20 Q_e p_e - 20 Q_e b_e + 3 Q_e^2  p_e^2
+ 9 Q_e^2  \alpha_e^2  + 60 e_1 - 18 Q_e \alpha_e e_1 
\cr& 
+ 6 Q_e^2 a_e \alpha_e + 6 Q_e^2 b_e \alpha_e + 6 Q_e^2 q_e a_e 
- 6 Q_e q_e e_1 - 3 Q_e^2 q_e^2
\cr&
+ 12 \alpha_d^2  + 18 e_1 \alpha_d + 304)/(12(4 + \alpha_d))
\cr}}

We can thus achieve cancellation of mixed anomalies while still retaining 
considerable freedom in the leptonic sector. 
Among the possible textures for $Y_e$ we have 
\eqn\leptexts{
Y_e^{I} \sim \lambda^{\alpha_e}
\pmatrix{\lambda^{4}&\lambda^{3}&\lambda^{3}\cr
\lambda^{3}&\lambda^{2}&\lambda^{2}\cr\lambda&1&1},
Y_e^{II}\sim \lambda^{\alpha_e}
\pmatrix{\lambda^{4}&\lambda^{4}&\lambda^{2}\cr
\lambda^{2}&\lambda^{2}&1\cr\lambda^2&\lambda^2&1},
Y_e^{III}\sim \lambda^{\alpha_e}
\pmatrix{\lambda^{4}&\lambda^{2}&\lambda\cr
\lambda^{4}&\lambda^{2}&\lambda\cr\lambda^3&\lambda&1}.}
All these textures correspond to the 
hierarchy $m_{\tau}:m_{\mu}:m_e = 1:\lambda^2:\lambda^4$.
From Eq.~\qecons{b}\ we see that for this class of textures 
\eqn\qqeres{Q_e = \frakk{2\alpha_d}{3(\alpha_e+2)} 
\quad\hbox{in the FI case}}
or
\eqn\qqeresr{Q_e = \frakk{2(3+\alpha_d)}{3(\alpha_e+2)} 
\quad\hbox{in the $\Rcal$ case,}}
so that, for 
example, in the FI case with with $\alpha_d = \alpha_e = 2$ we have 
$Q_e = 1/3$. Notice that in the $\Rcal$  case we can have $Q_e = 1$, 
if $\alpha_d = 3\alpha_e/2$.

\newsec{DD Textures}

In this section we consider an alternative texture solution, as described in 
Ref.~\jjwb\ (see also Ref.~\ref\EverettUP{
L.L.~Everett, G.L.~Kane and S.F.~King,
JHEP 0008  (2000) 012
}
This takes the form:

\eqn\textsdem{
Y_u \sim \pmatrix{\lambda^{8}&\lambda^{4}&1\cr
\lambda^{8}&\lambda^{4}&1\cr
\lambda^{8}&\lambda^{4}&1}, 
\quad
Y_d, Y_e \sim \lambda^{\alpha_{d,e}}
\pmatrix{\lambda^{4}&\lambda^{2}&1\cr
\lambda^{4}&\lambda^{2}&1\cr
\lambda^{4}&\lambda^{2}&1}.}
We will term these textures ``Doublet Democracy'' because (assuming 
as before a specific $\theta$-field is responsible for each texture)
it corresponds to generation-independent charges for quark and 
lepton doublets. Unlike the Wolfenstein case, the above textures 
do not lead naturally to a Wolfenstein texture for CKM;
in fact the entries in $U_{u,d}$, and hence CKM,  are generically of $O(1)$.
However, if we suppose that in fact
\eqn\qyuks{
Y_u \propto \pmatrix{a_u\lambda^8&d_u\lambda^4&1 + O(\lambda^2)\cr
b_u\lambda^8&e_u\lambda^4&1 + O(\lambda^2)\cr
c_u\lambda^8&f_u\lambda^4&1 + O(\lambda^2)}
\quad\hbox{and}\quad
Y_d \propto \pmatrix{a_d\lambda^4&d_d\lambda^2&1 + O(\lambda^2)\cr
b_d\lambda^4&e_d\lambda^2&1
+ O(\lambda^2)\cr c_d\lambda^4&f_d\lambda^2&1 + O(\lambda^2)\cr}}
(in other words that the unsuppressed Yukawa couplings are approximately 
the same in both cases)
then we obtain for the CKM matrix the texture
\eqn\ckmtext{CKM_{DD} \sim  \pmatrix{1&1&\lambda^2\cr
1&1&\lambda^2\cr\lambda^2&\lambda^2&1}
}
which is not of the form of the standard Wolfenstein parametrisation,
Eq.~\ckmwolf.
It does, however,  reproduce the most significant feature, which is
the smallness
of the couplings to the third generation. We obtain $CKM_{DD}$ 
although the entries in $U_{u,d}$ are generically still of $O(1)$, via 
a cancellation between $U_u$ and $U_d$. This observation will be important 
when  we come to consider neutrino masses in this scenario.

Unlike the Wolfenstein case, these textures do not dictate the 
value of $Q_d$.  
The anomaly constraints become (in Case~(FI)): 
\eqna\quafi$$\eqalignno{
A_3 &=                      12 + 3 \alpha_d Q_d + 6 Q_d - 3 \Delta = 0 
& \quafi a\cr
A_2 &=  -2 \Delta - 6 h_2 + 72 - 9 u_1 + 12 Q_e + 3 Q_e \alpha_e - 3 e_1
= 0 & \quafi b\cr
A_1 &= -120 + 6 \alpha_d Q_d + 12 Q_d 
+ 27 u_1 + 18 h_2 + 9 Q_e \alpha_e + 9 e_1 - 12 \Delta = 0 & \quafi c\cr
A_Q &= -48 u_1 - 144 h_2 - 96 Q_d - 48 \alpha_d Q_d 
- 28 Q_e^2  - 24 Q_e^2  \alpha_e + 12 Q_e e_1
 \cr& - 3 Q_e^2  \alpha_e^2  
+ 18 u_1 h_2 + 6 Q_e \alpha_e e_1 + 6 \alpha_d Q_d u_1
 + 12 \alpha_d Q_d h_2 + 48 \Delta  
+ 12 h_2^2 \cr& + 3 \alpha_d^2 Q_d^2  + 12 \alpha_d Q_d^2
 + 20 Q_d^2  + 12 Q_d u_1 + 24 Q_d h_2 + 224 + 24 Q_e \Delta  \cr& 
- 24 Q_e h_2 - 6 e_1 \Delta + 6 e_1 h_2 - \Delta^2  - 4 \Delta h_2 
- 12 Q_d \Delta - 6 u_1 \Delta\cr&
 + 6 Q_e \alpha_e \Delta - 6 Q_e \alpha_e h_2 - 6 \alpha_d Q_d \Delta
= 0 & \quafi d\cr}$$

and in Case~(${\cal R})$:
\eqna\quar$$\eqalignno{
A_3 &=  12 + 3 \alpha_d Q_d + 6 Q_d - 3 \Delta = 0 &\quar a\cr
A_2 &= 64 - 2 \Delta - 6 h_2 - 9 u_1 + 12 Q_e + 3 Q_e \alpha_e - 3 e_1
= 0 &\quar b\cr
A_1 &= -132 + 6 \alpha_d Q_d + 12 Q_d - 12 \Delta + 18 h_2 + 27 u_1 
+ 9 Q_e \alpha_e + 9 e_1= 0 &\quar c\cr
A_Q &= 176 + 6 Q_e \alpha_e e_1 + 42 \Delta - 132 h_2 - 42 u_1 
- 96 Q_d - 48 \alpha_d Q_d + 6 Q_e \alpha_e \cr& 
- 28 Q_e^2  - 24 Q_e^2  \alpha_e + 12 Q_e e_1 + 24 Q_e \Delta
- 24 Q_e h_2 - 3 Q_e^2  \alpha_e^2  
- 6 e_1 \Delta \cr& + 6 e_1 h_2 - \Delta^2
- 4 \Delta h_2 + 24 Q_e - 6 e_1 + 6 Q_e \alpha_e \Delta 
- 6 Q_e \alpha_e h_2 + 12 Q_d u_1 \cr& + 24 Q_d h_2 - 12 Q_d \Delta  
- 6 u_1 \Delta + 6 \alpha_d Q_d u_1
+ 12 \alpha_d Q_d h_2 - 6 \alpha_d Q_d \Delta \cr&  
+20Q_d^2+12h_2^2+18u_1h_2+3Q_d^2\alpha_d^2+12Q_d^2\alpha_d = 0. &\quar d\cr}$$

Solving Eqs.~\quafi{a-c}, we obtain
\eqna\quafisol$$\eqalignno{
Q_d &= \frakk{\Delta-4}{\alpha_d+2} & \quafisol a\cr
Q_e &= \frakk{2 (\Delta-6)}{3(2+\alpha_e)} & \quafisol b\cr
u_1 &= -2\Delta/9-2h_2/3+8+4Q_e/3+Q_e\alpha_e/3-e_1/3\cr
& = \frakk{4 \Delta-12 h_2-6 h_2 \alpha_e+96+60 \alpha_e-6 e_1-3 e_1 \alpha_e}
{9(2+\alpha_e)}.
& \quafisol c\cr
}$$
It would clearly be desirable to have $Q_d=Q_e=1$, since then we could
have a single $\theta$-charge only. 
However, imposing this would lead to $\alpha_d=\frak32(\alpha_e+2)$
which would not accord with the hierarchy of masses between leptons and 
quarks. If we 
instead  make the simplifying
assumption $\alpha_d=\alpha_e=0$, we find $Q_d=-2+\frak12\Delta$, 
$Q_e=-2+\frak13\Delta$. We also find 
\eqn\quadf{
A_Q = 
\Delta\left[\frak{26}{9}\Delta +4h_2-2e_1-128/3\right]
.}
We are clearly led to $\Delta=0$ since this gives $A_Q=0$ and also 
$Q_d = Q_e$, so that $\theta_e$ may be identified with $\theta_d$.  
Moreover, the fact that the $\theta_u$ 
and $\theta_d$ charges have 
{\it opposite\/} signs makes this assignment {\it natural}, 
in the sense  that we do not have to forbid higher dimensional terms 
involving powers of 
$\theta_{d,e}$ from contributing to $Y_u$.
It is this case we will 
concentrate on later. (After an exhaustive search we have been unable to find
any other ``natural'' $\theta$-charge assignments; it is worth mentioning
that simply requiring $Q_d=Q_e$ and $\alpha_d=\alpha_e$ leads inevitably to
the solution we have described.)  

Correspondingly from Eqs.~\quar{a-c}
we obtain instead (for Case~(${\cal R}$))
\eqna\quarsol$$\eqalignno{
Q_d &= \frakk{\Delta - 4}{\alpha_d+2}& \quarsol a\cr
Q_e &= \frakk{2 (\Delta-3)}{3(2+\alpha_e)} & \quarsol b\cr
u_1 &= 64/9-2\Delta/9-2h_2/3+4Q_e/3+Q_e \alpha_e/3-e_1/3\cr
&= \frakk{104+58 \alpha_e+4 \Delta-12 h_2-6 h_2 \alpha_e 
-6 e_1-3 e_1 \alpha_e}{9(2+\alpha_e)}.
& \quarsol c\cr
}$$
Interestingly in this case we {\it can\/} achieve 
$Q_d=Q_e=Q_u = 1$, as follows.  From Eq.~\quarsol{a,b}\ we have 
at once that $\alpha_d=\frak32\alpha_e$, which suggests
setting $\alpha_d=\alpha_e=0$ to avoid inverting the hierarchy between 
lepton and quark masses; we then have  $\Delta=6$. 
Now for $\alpha_d = \alpha_e =0$ we obtain
\eqn\quadr{
A_Q = 
\Delta\left[\frak{26}{9}\Delta +4h_2-2e_1-34\right]
-8h_2+4e_1 +44/3}
so that for $\Delta = 6$  
the constraint $A_Q = 0$ becomes
$2h_2 - e_1 = 32/3$. (Once again it turns out that this is the only 
solution even if we don't impose $\alpha_d=\alpha_e=0$ from the outset.)
Imposing the above constraint on $e_1$, we then have leptonic 
charges $L_i, e_1,e_2,e_3 = \frak{23}{3} -h_2, 2h_2 - \frak{32}{3}, 
2h_2 - \frak{38}{3}, 2h_2 - \frak{44}{3}$. Then the loop 
unsuppressed term from Eq.~\Rsol\ will be positive for each 
of the set $L_i, e_j$ if $m_0^2 < 0$ and 
$ 8 > h_2 > \frak{43}{6}$, which 
means the term will in each case help us eliminate the tachyonic slepton. 
We will analyse this case later;
unfortunately it runs into difficulties 
because it leads to 
a comparatively light charged Higgs mass. 
The alternative scenario which we described as natural in the FI case
was to arrange that $Q_d = Q_e < 0$. It is easy to demonstrate from 
Eqs.~\quarsol{a,b} that this is incompatible with $\alpha_d \leq \alpha_e$,
and moreover the alternative natural solutions
$Q_d = Q _u = 1$, $Q_e < 0$, or $Q_e = Q _u = 1$, $Q_d < 0$ are also
impossible;
since $Q_d = 1$ gives at once $\Delta \geq 6$ and hence $Q_e\ge 0$,
and $Q_e=1$ gives $\Delta\ge9$ and hence $Q_d\ge0$.

An alternative which we will consider later is
$\Delta = \alpha_d = \alpha_e = 0$, leading to $Q_d = -2$, $Q_e = -1$.  
This is clearly less satisfactory since we now must suppose that the 
physics responsible for the higher dimension terms does not permit 
$\theta_d$ to couple to the leptons. 

\newsec{Choice of Textures}

In the previous two sections we have exhibited two distinct choices of
Yukawa  texture and showed how they both could arise from $U'_1$ (or
$U^{\Rcal}_1$)  charge assignments compatible with mixed-anomaly
cancellation. 

The Wolfenstein texture has the advantage of explaining in a completely 
natural way the origin of the CKM matrix; however the DD texture 
has an overriding advantage which is specific to our AMSB scenario.     
This advantage derives from the fact that for the corresponding textures 
$Y_{u,d,e}$ shown in Eq.~\textsdem, the right-handed diagonalisation 
matrices $V_{u,d,e}$  are close to the unit matrix. Specifically, 
\eqn\wtexts{
V_u \sim \pmatrix{1&\lambda^{4}&\lambda^{8}\cr
\lambda^{4}&1&\lambda^{4}\cr
\lambda^{8}&\lambda^{4}&1}, 
\quad
V_d, V_e \sim
\pmatrix{1&\lambda^{2}&\lambda^{4}\cr
\lambda^{2}&1&\lambda^{2}\cr
\lambda^{4}&\lambda^{2}&1}.}

The significance of this becomes apparent when we consider the effect 
of rotating to the quark/lepton mass diagonal basis the fundamental relations 
Eq.~\massrgb\ or Eq.~\Rsol.
The AMSB   
contributions to the scalar masses are diagonalised to a good approximation
when we transform to
the  fermion mass-diagonal basis, as are the contributions proportional to 
$\gamma^i{}_j$ in Eq.~\Rsol. Clearly the danger lies in the terms linear in 
the $U'_1$ ( or $U_1^{\cal R}$) charges. However if we choose 
the DD textures, then on the one hand 
there is no problem with the LH squarks  and  sleptons, 
because of the universal doublet $U'_1$ charges; and on the other hand, 
the induced off-diagonal contributions to the RH squark  and  slepton
mass matrices are small because of Eq.~\wtexts\ above. 

For the rest of this paper we will concentrate on the DD textures. 

\newsec{Experimental Consequences}

The gaugino spectrum is to leading order independent of the  mechanisms
used here to resolve the slepton mass problem; and is characterised by  an
approximately degenerate triplet of light winos ($\Wtil^{\pm,0}$)
The neutral wino is, in a substantial region of parameter
space, the LSP; the resulting characteristic decay 
$\Wtil^{\pm}\to\Wtil^0\pi^0$ has been described in a number of papers. 

However the LSP can also be a  scalar neutrino, $\nutil_{l}$, 
in which case the dominant decay modes of $\Wtil^{\pm,0}$
will be $\Wtil^{\pm}\to\nutil_{l}l$  and 
$\Wtil^0\to \nutil_{l}\nu_{l}$ respectively (if the 
masses are ordered $\ltil > \Wtil^{\pm,0} > \nutil_{l}$) with 
the possibilities $\Wtil^{\pm}\to\ltil \nu_l, \Wtil^0\to \ltil l$
also available if $\Wtil^{\pm,0} > \ltil > \nutil_{l}$

The fact that $M_3$ and $M_2$ have opposite signs disfavours at first sight  
a \sic\ explanation of the well-known discrepancy 
between theory and experiment for  the anomalous magnetic moment 
of the muon, $a_{\mu}$. This is 
because if sign ($\mu M_2$) is chosen so as to create 
a positive $a_{\mu}^{\rm{SUSY}}$ then sign ($\mu M_3$) leads to 
constructive interference between various \sic\ contributions to
$B(b \to s\gamma)$, and consequent restrictions on the allowed 
parameter space\foot{In this context {\it deflected\/} 
anomaly mediation\ref\RattazziQG{
R.~Rattazzi, A.~Strumia and J.D.~Wells,
\npb 576  (2000) 3
}\
is worthy of consideration\ref\AbeEQ{
N.~Abe and M.~Endo,
hep-ph/0212002
}.}.
It is   worth noting, however, that gloomy conclusions here have
generally  been reached in the context of  the mAMSB model,
Eq.~\massrgnot, and the issue is therefore perhaps worth  revisiting, 
because the nature of our solution to the tachyonic slepton
problem means that the squark/slepton mass difference  is  typically
much  higher in our case,  so we might hope (by choosing positive $\mu
M_2$,  and arranging for heavy squarks) 
to find a negligible (even if constructive) $B(b
\to s\gamma)$  contribution from squark loops.

We must be careful,
however, of the charged Higgs contribution to $B(b \to s\gamma)$, which
can be quite large, is independent of squark masses,  and {\it adds\/}
to the SM contribution. Ignoring other \sic\ contributions, we estimate
that a limit  $m_{H^{\pm}} > 400\GeV$   is required. This limit provides
a useful constraint on our final choice of charge assignments, via  mass
sum rules, which we will discuss below.

\subsec{The FI case}
In section~6 we showed how with  
$\Delta = \alpha_d = \alpha_e = 0$ anomaly cancellation led to 
the economical and natural result 
$Q_d = Q_e = -2$ so that two  $\theta$-fields
$\theta_u$, $\theta_d$  suffice to generate the quark and lepton masses.
Moreover, as mentioned earlier, there is no need to invoke any further
discrete symmetry to forbid $\theta_d$ from contributing to $Y_u$ 
since $\theta_u$, $\theta_d$ have opposite charges.
If one assumed that only these two fields received vevs at the 
$M_{\theta}$-scale, and if $U'_1$ were gauged, then there would  be  
a D-flat direction corresponding to 
$\vev{\theta_u} = \sqrt{2}\vev{\theta_d}$ leading to two distinct 
$\lambda$-parameters, with $\lambda_u = \sqrt{2}\lambda_d$. However, we
retain an agnostic attitude to the gauging of the $U'_1$ and we choose
correspondingly  
to stick with the basic assumption that there is a universal 
$\lambda \approx 0.22$ for all the Yukawa matrices.

In Ref.~\jjwb\ we presented a preliminary analysis of this scenario. 
The main distinguishing feature of the sparticle spectrum is the 
large splitting among right-handed fields caused by 
the generation-dependent charge assignments. This is in 
complete contrast to the constrained MSSM (CMSSM), where, for example, 
$\dtil_R,\stil_R$ are almost degenerate as are $\dtil_L,\stil_L$.
Here while $\dtil_L,\stil_L$ remain degenerate, $\dtil_R$ and 
$\stil_R$ may differ in mass by a factor of two or more. 

In order to pin down an appropriate set of $U'_1$ charge assignments 
(and also to explore what features of the outcome are independent 
of these assignments) it is useful to begin by introducing some sum rules. 
It is easy to show that in Case~(FI) we have, after imposing 
the anomaly constraints Eqs.~\quafisol{}, 
\eqna\sumrules$$\eqalignno{
\Tr\left(m_L^2+3m_Q^2\right) &= 
\Tr\left(m_L^2+3m_{Q}^2\right)|_{\rm AMSB} -\Delta\zeta', & \sumrules a\cr
\Tr\left(m_{u^c}^2+m_{d^c}^2+2m_Q^2\right) 
&= \Tr\left( m_{u^c}^2+m_{d^c}^2+2m_Q^2\right)|_{\rm AMSB},& \sumrules b\cr
\Tr\left(m_{u^c}^2+m_{e^c}^2-2m_Q^2\right) 
&= \Tr\left(m_{u^c}^2+m_{e^c}^2-2m_Q^2\right)|_{\rm AMSB}, & \sumrules c\cr}
$$
where the masses on the RHS correspond to pure AMSB contributions, 
i.e. they are calculable from Eq.~\amsbsome{c}, which (apart from the overall 
scale $M_0$), depends only on the unbroken theory. We hence obtain 
sum rules for the particle masses, for example from Eq.~\sumrules{b}\ we have 
that: 
\eqn\trsumrules{
\sum_{\qtil} m^2_{\qtil} = 2m_t^2 
+ j_1(\tan\beta) m_{\gtil}^2 
}
where  the sum includes all twelve squarks, and we have neglected 
quark masses apart from $m_t$. The function $j_1(\tan\beta)$ is a 
slowly varying function of $\tan\beta$, given to 
a good approximation by $j_1 = 9.84 - 0.013\tan\beta$ for 
$5 < \tan\beta < 40$. For $\tan\beta < 5$ and $\tan\beta > 40$, $j$ 
increases, for example $j_1(2) = 9.91$ and $j_1(60) = 10.02$.
This sum rule is very robust, being 
independent of $\Delta$ and any feature of the charge assignments. 
Note how it clearly distinguishes the FI case from the 
mAMSB variant defined by Eq.~\massrgnot, 
which would lead to an additional 
term $12m_0^2$ on the RHS of Eq.~\trsumrules, which would also hold only 
at high energies because of the loss of RG invariance.

Adding Eqs.~\sumrules{a,c}we obtain 
\eqn\trsumrulesb{
\sum_{\ttil,\ctil,\util} m^2_{\qtil} +\sum_{\tautil,\mutil,\etil} m^2_{\ltil} 
= 2m_t^2 
+ j_2(\tan\beta)  m_{\gtil}^2 -\Delta\zeta' 
}
so for the class of charge assignments such that $\Delta = 0$ (corresponding 
to an allowed $\mu$-term $\mu H_1 H_2$) we have another sum rule. The 
function $j_2(\tan\beta) \approx 4.6$ is again 
insensitive to $\tan\beta$.
 
There is a further sum rule involving the CP odd Higgs. Using the tree 
minimisation conditions we obtain
\eqn\hggss{
m_A^2 = (m_2^2 - m_1^2)\sec 2\beta - M_Z^2}
whence 
\eqn\hggsr{
m_A^2
= \sec 2\beta\left(\frakk{2}{3}\sum_{\tautil,\mutil,\etil} m^2_{\ltil} 
+ j_3(\tan\beta) m_{\gtil}^2 + (8-\Delta/3)\zeta'\right)}
where $j_3$ 
increases from  $j_3(5) \approx -0.45$ to 
$j_3(40) \approx 0.01$ .

This sum rule is particularly useful in the $B(b \to s\gamma)$ context,
because $m_A$ is linked to $m_{H^{\pm}}$ via 
the tree relation $m_{H^{\pm}}^2 = m_A^2 + M_W^2$, 
and as we described above, we 
want  to ensure  $m_{H^{\pm}} > 400 \GeV$. We must also ensure 
that the tachyonic mass problem is solved. For 
$\Delta = \alpha_d = \alpha_e = 0$ so that 
from Eq.~\quafisol{c} we have $3u_1 = 16-2h_2-e_1$, 
we have $U'_1$ charge assignments  
as shown in Table~1, 
\vskip3em
\vbox{
\begintable
 Q_i|u_2 | u_3  | d_1 | d_2  |d_3 
\cr
8 -u_1-h_2|u_1-4  | u_1-8
|2h_2+u_1-16|2h_2+u_1-12 |u_1+2h_2-8
\endtable
\bigskip
\begintable
L_i |  e_1 | e_2 | e_3 
\cr
3u_1+3h_2-24|16-2h_2 -3u_1 |20-2h_2-3u_1 
|24-2h_2-3u_1 
\endtable
\inparg
{\noindent\hskip 2cm {\it Table~1:\/} The $U'_1$-charges}
\bigskip \outparg}
\noindent
where we have written the charges in terms of $u_1$ instead of $e_1$ 
using Eq.~\quafisol{c}. 
It is then easy to show that 
with the charge assignments shown in Table~1, 
there exists some range of $\zeta, \zeta'$ leading 
to positive  FI contributions for both $m^2_{e^c}$ and $m^2_L$ 
if and only if 
\eqn\posi{3u_1+4h_2 < 24, \quad\hbox{if}\quad \zeta' < 0,}
or 
\eqn\posib{3u_1 + 4h_2 > 32 \quad\hbox{if }\quad \zeta' > 0.}
Now in Ref.~\jjwb\ we chose $h_2 = 12$ and $u_1 = -7/2$, which 
evidently satisfies Eq.~\posib. Throughout the corresponding 
allowed region in the $\zeta,\zeta'$ plane, however,  this gives rise to 
an unacceptably light $H^{\pm}$  mass from the point of view described 
above. The reason is easy to see from Eq.~\hggsr; for 
$\tan\beta >1$ we have $\sec 2\beta < 0$, and so the last term in this 
equation reduces $m_A^2$ (and hence $m_{H^{\pm}}$) if $\zeta' > 0$. 
If, however, we choose, for example, $h_2 = 1$ and  $u_1 = 1/2$ 
(corresponding to $e_1 = 25/2$), then 
we have instead $\zeta' <  0$
and duly obtain a spectrum with a significantly heavier $H^{\pm}$.

For these charge assignments, $\tan\beta=5$, 
$\mu > 0$ and $M_0=40\TeV$, we show in Figure~1 
the triangular region in the
$\zeta_{1,2}$ plane which corresponds to an acceptable vacuum. 
The LSP can be the neutral wino,
or the $\nutil_{\tau}$; for alternative charge assignments 
(such as those employed in Ref.~\jjwb) the LSP can be a charged lepton,
but we find that the constraint of a heavier $H^{\pm}$ that we favour here 
excludes this possibility.    
\smallskip
\epsfysize= 4.0in
\centerline{\epsfbox{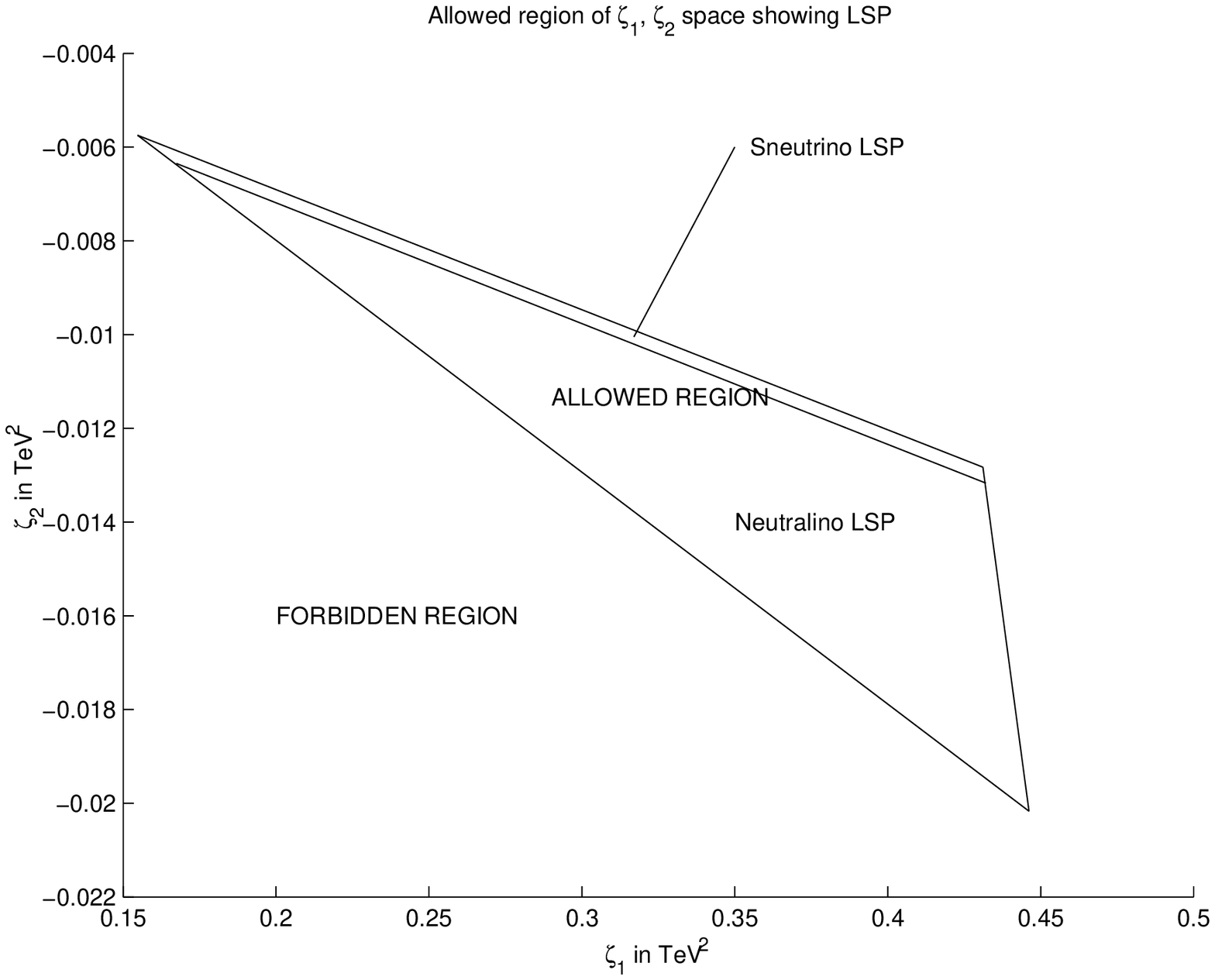}}
\inparg
{\it \noindent Fig.1:
Allowed values of $\zeta_{1,2}$ for $\tan \beta=5$,
$m_0=40\TeV$ and $\mu > 0$.
}
\smallskip
\outparg

As $\tan\beta$ is increased the allowed region shrinks, becoming 
very small for $\tan\beta > 20$.
In Table~2 we give representative spectra
for a point from each of the two allowed regions in Fig.~1.

If we consider  the muon anomalous magnetic moment for the 
case, for example, of the second column of Table~2, then 
we obtain\ref\amuth{T.~Moroi, \prd 53 (1996) 6565 
[Erratum-{\it ibid\/} \prd 56 (1997) 4424]\semi
S.P.~Martin and J.D.~Wells,
\prd 64   (2001) 035003}
\eqn\anommu{a_{\mu}^{\hbox{SUSY}}\approx 25\times 10^{-10}.}
This result (the dominant contribution to  which comes from the 
charged wino/Higgsino diagram) 
is in the right region to explain the difference between 
the recent Brookhaven E821 result
\ref\BennettJB{
G.W.~Bennett {\it et al.}  [Muon g-2 Collaboration],
\prl 89  (2002) 101804
[Erratum- {\it ibid.\/} 89  (2002) 129903]
}
and the SM prediction\foot{We use the $e^+ e^-$ result from 
Ref.~\ref\DavierDY{
M.~Davier, S.~Eidelman, A.~Hocker and Z.~Zhang,
hep-ph/0208177.
};  see also 
Ref.~\ref\amuothers{
S.~Narison,
\plb 513 (2001) 53 
[Erratum-{\it ibid.\/} \plb 526  (2002) 414]\semi
J.F.~De Troconiz and F.J.~Yndurain,
\prd 65 (2002) 093001\semi
K.~Hagiwara, A.D.~Martin, D.~Nomura and T.~Teubner,
hep-ph/0209187
}}
\eqn\amudis{\delta a_{\mu} = 33.9(11.2) \times 10^{-10}}
Thus (by choosing $\mu > 0$) we are indeed able to generate a significant 
positive contribution to $a_{\mu}$ while simultaneously suppressing the 
contribution to $B(b \to s\gamma)$ thanks to the large 
squark and Higgs masses. 

\vskip3em
\vbox{
\begintable
 \tan\beta (\hbox{sign}~\mu_s) | 5(+)  |5(+)  \cr
  \zeta_1 (\TeV)^2 |0.3 | 0.325 \cr
 \zeta_2 (\TeV)^2 |-0.01 | -0.01 \cr
 M_0\TeV |  40 |40  \cr
\hbox{mod}(\mu) \TeV |0.677  |0.667  \cr
  \ttil_{1,2} | 863,606 |864,594\cr
\ctil_{L,R} |931,852| 933,842\cr
 \util_{L,R} |931,828 |933,818\cr
 \btil_{1,2} | 825,1024  |827,1028\cr
\stil_{L,R} |934,1047| 936,1051\cr
 \dtil_{L,R} |934,1066|936,1069\cr
 \tautil_{1,2} |  150,268 | 101,311\cr
\mutil_{L,R} |152,335| 104,370\cr
 \etil_{L,R} |152,390|104,421\cr
\nutil_{\tau}| 129|64\cr
 \nutil_{\mu, e} |130|66\cr
 h | 117| 117 \cr
 H | 539 | 514 \cr
 A | 538| 512\cr
 H^{\pm} |544 |519\cr
 \chitil^{\pm}_1 | 104 |104 \cr
 \chitil^{\pm}_2 | 681 |671 \cr
 \chitil_1 |103| 103 \cr
 \chitil_2 | 367 |367 \cr
 \chitil_3 | 680 | 670\cr
 \chitil_4 |689| 680\cr
 {\tilde g} | 1008 | 1008\endtable}
\inparg
{\noindent\hskip 2cm {\it Table~2:\/} The sparticle masses 
(in $\GeV$) for the FI $U_1'$ case
\bigskip} 
\outparg
\subsec{The ${\cal R}$ case}

Here also we will be guided by the Higgs mass sum rule. 
The sum rule analogous to Eq.~\trsumrules\ is 
\eqn\trsumrulesR{
\sum_{\qtil} m^2_{\qtil} = 2m_t^2
+ j_1 (\tan\beta) m_{\gtil}^2 + k^{\Rcal}_1 (\tan\beta) m_0^2
}
where $k^{\Rcal}_1 (5)\approx 2.7$ 
and is   likewise slowly varying over 
a wide range of $\tan\beta$. Thus there is explicit dependence on the 
mass scale $m_0$; this term is small compared to the $m_{\gtil}$
term, however. 

Similarly  we have a sum rule like Eq.~\trsumrulesb:
\eqn\trsumrulesbR{
\sum_{\ttil,\ctil,\util} m^2_{\qtil} +\sum_{\tautil,\mutil,\etil} m^2_{\ltil}
= 2m_t^2
+j_2(\tan\beta) m_{\gtil}^2 + 
(\frak{3}{2}\Delta- k^{\Rcal}_2 (\tan\beta))m_0^2}
where $k^{\Rcal}_2 (5)\approx 6.2$.
Finally the analogue of the Higgs mass sum rule, Eq.~\hggsr\ is 
\eqn\hggsrR{
m_A^2 =
\sec 2\beta\left(\frakk{2}{3}\sum_{\tautil,\mutil,\etil} m^2_{\ltil} 
+ j_3 (\tan\beta) m_{\gtil}^2 + {\cal C}m_0^2\right)}
where   
\eqn\ccaldef{
{\cal C} = \gamma_{H_2} - \gamma_{H_1} - \frak{2}{3}\Tr(\gamma_L + \gamma_{e^c})
+\Ccal_q}
and
\eqn\Cqdef{\Ccal_q = 2 + \sum_{i=1}^3 (L_i + e_i) - \frak{3}{2}(h_2 - h_1).} 
Here there is a degree of cancellation between the first two terms on the 
RHS of Eq.~\hggsrR\ and hence 
the sign and magnitude of $\Ccal_q$ becomes important.
For the case $\Delta = 6$, $Q_u = Q_d = Q_e =1$ described at the end of 
section~6, we find (independent of $h_2$) that $\Ccal_q = -4$, so that 
since we needed $m_0^2 < 0$  to resolve the tachyonic slepton problem 
we can anticipate light $m_A, m_{H^{\pm}}$. 
We indeed find that for $M_0 = 40\TeV$, 
$m_{H^{\pm}} \approx 160-250\GeV$, and that this continues to hold even 
if we raise $M_0$ to $80\TeV$, giving squark masses in the region of $2\TeV$
and $|\mu| \approx 1.2\TeV$, which is at if not beyond the limit 
of acceptable fine tuning for the Higgs minimisation. Thus this scenario, 
while ideal in terms of the $\theta$-charges, is, we believe, ruled out.  

An alternative for which we will again provide detailed results is
$\Delta = \alpha_d = \alpha_e = 0$, leading to $Q_d = -2$, $Q_e = -1$. 
Unfortunately as we already described, 
here we have to forbid $\theta_d$ from coupling to the leptons
in order to prevent it from giving unwanted contributions to the textures for 
$Y_e$. 
We now find that 
$\Ccal_q = -7$. However, this time solving  $A_Q = 0$ gives 
$e_1=2h_2 - 11/3$, whereupon the leptonic charges are 
$L_i, e_1,e_2,e_3 = -\frak{4}{3} -h_2, 2h_2 - \frak{11}{3},
2h_2 - \frak{5}{3}, 2h_2 + \frak{1}{3}$. Then the loop
unsuppressed term from Eq.~\Rsol\ will be positive for each
of the set $L_i, e_j$ if $m_0^2 > 0$ and $-1 < h_2 < -1/3$.
So in this case we can have  $m_0^2 > 0$ with $\Ccal_q < 0$ and hence 
a {\it positive\/} $\Ccal$-contribution to $m_A^2$ in Eq.~\hggsrR\ leading 
to a larger $m_{H^{\pm}}$. 
For these charge assignments, $h_2 = -2/3$ and  $\mu > 0$  we show
in Figure~2 the  range of values of $M_0$ and $m_0$ that lead to an
acceptable  vacuum and sparticle spectrum, for both $\tan\beta = 5$ and 
$\tan\beta = 10$.  In both cases the  LSP is always the $\nutil_{\tau}$,
which is disfavoured as a dark matter candidate. 

\smallskip
\epsfysize= 3.7in
\centerline{\epsfbox{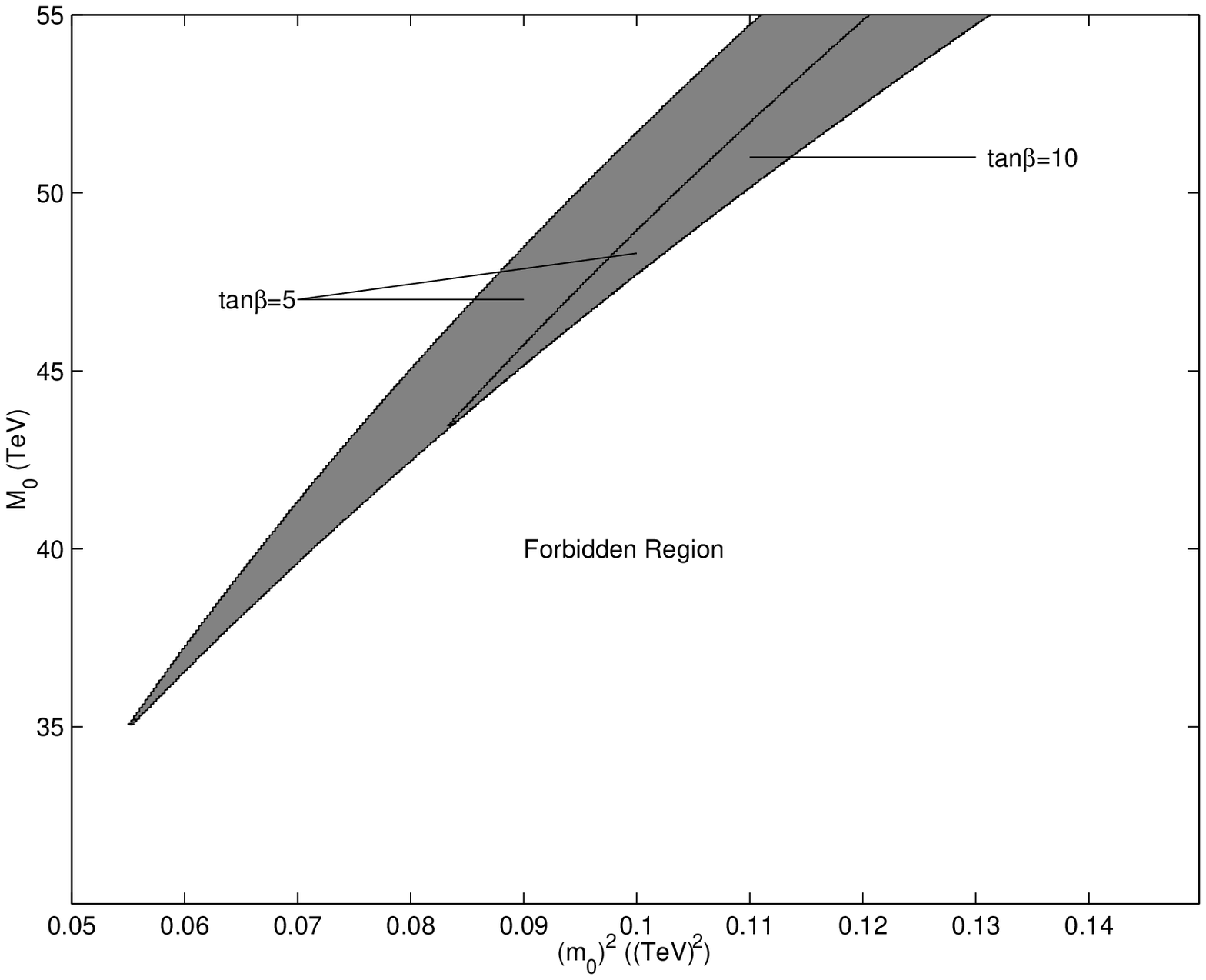}}
\inparg
{\it \noindent Fig.2:
Range of allowed values of $M_0$ and $m_0$ for the $U^{\Rcal}_1$ case}
\smallskip
\outparg

In Table~3 we give representative spectra
for two points from the allowed region in Fig.~2, corresponding 
to $\tan\beta = 5$ and $\tan\beta = 10$ respectively. 
\vskip3em
\vbox{
\begintable
 \tan\beta (\hbox{sign}~\mu_s) | 5(+)  |10(+)  \cr
 M_0 (\TeV) |  40 |50 \cr
 m_0^2 (\TeV^2) |0.07 |0.104 \cr
\hbox{mod}(\mu) \TeV |0.753  |0.929  \cr
  \ttil_{1,2} | 872,674 |1059,827\cr
\ctil_{L,R} |932,667| 1152,836\cr
 \util_{L,R} |932,159 |1152,274\cr
 \btil_{1,2} | 826,1004  |1011,1228\cr
\stil_{L,R} |935,1198|1155,1473\cr
 \dtil_{L,R} |935,1362|1155,1672\cr
 \tautil_{1,2} |  108,218 | 101,261\cr
\mutil_{L,R} |114,507| 128,615\cr
 \etil_{L,R} |114,683|128,831\cr
\nutil_{\tau}| 79|89\cr
 \nutil_{\mu, e} |81|97\cr
 h | 117| 124 \cr
 H | 664 |781 \cr
 A | 663| 781\cr
 H^{\pm} |668 |785\cr
 \chitil^{\pm}_1 | 105 |134 \cr
 \chitil^{\pm}_2 |757|931\cr
 \chitil_1 |104| 133 \cr
 \chitil_2 | 368 |463 \cr
 \chitil_3 | 764 | 932\cr
 \chitil_4 |756| 937\cr
 {\tilde g} | 1008 | 1246 \endtable}

\inparg 
{\noindent\hskip 2cm {\it Table~3:\/} The sparticle masses    
(in $\GeV$) for the $U^{\Rcal}_1$ case       
\medskip} 
\outparg  

Here a significant constraint on the allowed parameter space 
is provided by the $\util_R$ mass, which tends to be lighter than the 
other squarks. Otherwise the spectrum is similar to that obtained 
for the $\nutil_{\tau}$ region in the FI case.

\newsec{R-parity violation}

It is well-known that imposing gauge invariance alone does not 
forbid the addition of the following renormalisable
$R$-parity violating terms to the superpotential of the MSSM :
\eqn\Rviol{
W_R = \lambda_{ijk}L_i L_j e^c_k + \lambda'_{ijk}L_i Q_j d^c_k
+\lambda''_{ijk}u^c_id^c_j d^c_k + m^R_i L_i H_2.}
If all these terms are allowed then rapid proton decay results 
but if $\lambda''$ is forbidden then the limits on the remaining terms 
(which violate $L$ but not $B$) are less strict. 
Indeed the  cubic terms have been employed to provide an alternative 
explanation for the tachyonic slepton problem\ref\AllDed{
B.C.~Allanach and A.~Dedes,
JHEP 0006  (2000) 017}. 
Alternatively if only the quadratic term 
is permitted then this  results in interesting 
phenomenology\ref\campos{F.~de Campos et al, \npb   623 (2002) 47}; 
for example the mixing between neutral gauginos 
and neutrinos induces neutrino masses.

When  we generalise
the MSSM to the $\MSSMnu$ (Eq.~\supernu), 
i.e. if we consider the effective field theory at scales $P$ such that
$M_{\nu} < P < M_{\theta}$
\foot{In fact  the most natural assumption is $M_{\nu} = M_{\theta}$,
in which case we would also not have to be concerned with 
lepton flavour violation effects generated from $Y_{\nu}Y^{\dagger}_{\nu}$ contributions 
to the running of the slepton masses between $M_{\theta}$ and $M_{\nu}$; 
for a review and references see for example 
\ref\rnma{R.N.~Mohapatra, hep-ph/0211252}.}  
then
we have the further possible renormalisable $R$-parity violating terms
\eqn\rviolnu{
W_R^{\nu} = a_i \nu^c_i + b_i H_1 H_2 \nu^c_i + c_{ijk}\nu^c_i\nu^c_j\nu^c_k.}

Let us now ask the following question: can our $U'_1$ or 
$U^{\Rcal}_1$
be used to naturally forbid the appearance of some or all of 
the operators in Eqs.\Rviol, \rviolnu\  at the tree level or  
both at the tree level and via FN texture terms?

The answer to both these questions is in fact yes, 
there exist charge assignments which do precisely this. This is an attractive 
feature of this class of theories; the fact that a symmetry introduced 
to resolve the slepton mass problem can lead naturally to 
R-parity conservation is very economical. As an example of how this 
may be achieved consider the charge assignments that we studied in section~8 
in the FI case (from Eq.~\quafisol): $\Delta = 0, h_2 = 1, e_1 = 25/2$, with 
$\alpha_d = \alpha_e = 0$.  This corresponds to 
$Q_u = 1, Q_d = Q_e = -2$ so that manifestly only operators with integer 
charges can be generated with the possible $\theta$-charges. However 
we find that the set of $R$-parity violating operators of Eq.~\Rviol\ 
have charges $-27/2, -35/2, -37/2, -43/2, -51/2, -59/2$. 
Because these are all half-integral 
they cannot be generated by the existing $\theta$-charges and so $R$-parity
conservation is exact.   With regard to Eq.~\rviolnu, the outcome depends on 
the $\nu^c$ $U'_1$ charge assignments; we will reurn to this issue in the 
next section. 

In the $\Delta = \alpha_d = \alpha_e = 0$ (so that $Q_d=-2, Q_e = -1$), 
$\Rcal$-case analysed in Section~8.2, all the 
lepton number violating operators have fractional charges and 
are hence forbidden in the same manner, 
as are {\it tree\/} contributions to $\lambda''_{ijk}$. However if we allow  
FN couplings to the $\theta$-fields, we can in this case generate 
some of these  terms with powers of $\theta$ ranging from unity 
(for $\lambda''_{113}, \lambda''_{223}$) up to 
$\theta^7$ for $\lambda''_{312}$, if we allow any of $\theta_{u,d,e}$
in each case. 
In fact it is  $\lambda''_{112}$ and 
$\lambda''_{113}$ which are subject to the most 
strict experimental bounds (on double nucleon decay and $n-\nbar$ 
oscillations)\ref\DreinerUZ{
H.K.~Dreiner, hep-ph/9707435\semi
B.C.~Allanach, A.~Dedes and H.K.~Dreiner,
\prd 60 (1999) 075014}. 
These operators can be generated by $\theta_d^3$ and $\theta_d$
(or $\theta_e^6$ and $\theta_e^2$) respectively, and so 
we need to impose that these operators cannot be produced via the 
$\theta_{d,e}$ spurions.

\newsec{Neutrino Masses}

There is now substantial evidence for the existence of neutrino masses,
and also for a form of $U_{MNS}$ quite different from the CKM matrix. 
Such a difference is not surprising in the  context of the seesaw
model, since although $U_e$ is analogous to $U_{u,d}$, $U_{\nu}$ 
is quite different, involving as it does the singlet mass matrix 
$M_{\nu^c}$. This issue has been discussed at length in the literature; 
see for example the papers of King 
(Ref.~\ref\sfking{S.F.~King,
hep-ph/0208266
} and references therein).
Specific to the the DD scenario, however, there is a reason why 
we might expect a distinction between CKM and $U_{MNS}$. 
The small angles of CKM are produced 
by cancellation between $U_{u}$ and $U_{d}$ in Eq.~\ckmdef; the DD 
texture form produces texture suppression of the off-diagonal 
elements in $V_{u,d,e}$ but not in $U_{u,d,e}$. Therefore, in fact, 
(since $U_e$ has a similar form to $U_d$), 
we would anticipate (even if $U_{\nu}\approx 1$) a non-CKM form for 
$U_{MNS}$.

It is easy to provide an explicit realisation. Suppose that 
$m_{\nu}$ is to a good approximation diagonal, so that $U_{\nu} \approx 1$. 
This is possible in the context of the explicit texture-generated 
construction of Ref.~\jjwb, where we had 
\eqn\mdirac{m_D
 = \pmatrix{a_{\nu}\lambda^n&d_{\nu}\lambda^m\cr
b_{\nu}\lambda^n&e_{\nu}\lambda^m\cr c_{\nu}\lambda^n&f_{\nu}\lambda^m\cr},}
and 
\eqn\majmat{ M_{\nu^c} = \pmatrix{0&M^{\nu}_1\cr M^{\nu}_1&0}.}
If $a_{\nu}$, $d_{\nu}$ and 
$b_{\nu}f_{\nu}+ c_{\nu}e_{\nu}$ are small, then $m_{\nu}$ 
(which has one zero eigenvalue, because we have introduced only two 
right-handed neutrinos) is 
approximately  diagonal.
Now given a unitary matrix 
\eqn\Udef{U_e = \pmatrix{u_{11}
&u_{12}&u_{13}\cr u_{21}&u_{22}&u_{23}\cr u_{31}&u_{32}&u_{33}\cr},} 
then trivially the matrix
\eqn\MLdef{Y_e = \pmatrix{B\lambda^4 u_{11}
&A\lambda^2 u_{12}&u_{13}\cr B\lambda^4 u_{21}&A\lambda^2u_{22}&u_{23}\cr 
B\lambda^4 u_{31}&A\lambda^2u_{32}&u_{33}\cr}}
has the appropriate texture form (see Eq.~\textsdem) and 
is diagonalised by a left-handed transformation only:
\eqn\Ldiag{
U_e^T Y_e = \pmatrix{B\lambda^4
&0&0\cr 0&A\lambda^2&0\cr 0&0&1\cr}}

Thus if the neutrino mass matrix $m_{\nu}$ 
is to a good approximation diagonal 
and $Y_e$ is of the form above then we obtain 
$U_{MNS} = U^T_e$, a natural hierarchy for the charged lepton masses, and 
natural suppression of leptonic FCNCs. Clearly a suitable form for $U_e$ 
would be, for example (see Eq.~\numixa)
\eqn\Uposs{
U_e = 
\pmatrix{\frak{1}{\sqrt{2}}&\frak{1}{2}&\frak{1}{2}\cr
-\frak{1}{\sqrt{2}}&\frak{1}{2}&\frak{1}{2}
\cr 0 & -\frak{1}{\sqrt{2}}&\frak{1}{\sqrt{2}}\cr}}

The coefficients $u_{ij}$ depend on unknown physics, but this at 
least shows that a plausible $U_{MNS}$ is possible within our framework.
Finally we remark that if we choose (in Eq.~\mdirac) $n=2$ and $m=1$, 
then with the set of FI charge assignments that we favoured 
in previous sections we find charges $\theta_{\nu} = -37/3$ and 
$\nu^c = \pm 37/6$ which means that all the terms in Eq.~\rviolnu\ are,
like those in Eq.~\Rviol, forbidden at both tree and 
texture-generated level.

\newsec{Conclusions} 

We have given a detailed analysis of the constraints that follow from
imposing  mixed-anomaly cancellation on the $U'_1$  charge assignments
associated with Yukawa textures (both a $U'_1$ commuting with
supersymmetry  and a $U^{\Rcal}_1$).  The resulting texture patterns 
are of interest in their own right; however our specific interest is  in
application to the AMSB framework. Introducing the mixed-anomaly-free 
$U'_1$ allows the AMSB slepton mass problem to be solved while
maintaining RG invariance  and UV insensitivity.   In order to generate
Yukawa textures the $U'_1$ charge assignments must be  generation
dependent, which leads to potential FCNC problems. We have shown   that
a specific form for the textures solves this problem in a natural  way
without fine-tuning. The resulting spectrum patterns are clearly 
distinguished both from the CMSSM and the mAMSB by,  for example, large
squark and slepton mass splittings.   
Of the two $U_1$ scenarios we present, we favour the ordinary $U'_1$ 
case over the $U^{\Rcal}_1$ one. In the latter case although we 
were able to achieve the most economical $\theta$-charge structure 
($\theta_u = \theta_d = \theta_e = 1$), we find that in that case 
the charged Higgs mass is inevitably rather light; we present an alternative 
assignment ($\theta_u = 1, \theta_d = -2, \theta_e = -1$) which 
avoids this problem, but is unnatural in that higher dimension operators 
involving $\theta_d$ contributing to the lepton Yukawa 
matrix must be forbidden by fiat; also in this case the LSP is 
always a sneutrino, which is disfavoured as a dark matter candidate. 
Our FI case avoids all these criticisms. In addition all 
$R$-parity violating operators are naturally forbidden. 
 
Neutrino masses and mixings consistent with current observations  can be
accommodated within our framework. The matrix $U_e$ that rotates the 
left-handed charged leptons to the mass diagonal basis has generically  
large angles which can explain the difference between  the CKM matrix and
$U_{MNS}$, although  specific features of preferred patterns,  such as
near-vanishing of the $(e3)$ element of $U_{MNS}$, are not predicted. 

We believe that both the flavour-blind framework of Ref.~\jjfi\
and the texture based frameworks of this paper are more 
attractive possibilities than  mAMSB (and even arguably the CMSSM)
and consequently worthy of attention.

\appendix{A}{The AMSB solution}

Here we summarise the exact results for the soft \sy-breaking 
$\beta$-functions for a general $N=1$ \sic\ gauge theory with a simple
gauge group 
including 
the possibility of gauge singlets. We then give a set of results 
for the various soft parameters which form an exact RG trajectory, expressing 
them in terms of a single mass scale $M_0$ and the coupling constants of the
unbroken theory.  In 
Ref.~\ref\jjwa{I.~Jack, D.R.T.~Jones and R.~Wild, \plb 509 (2001) 131}, 
we distinguished results according to whether the auxiliary $F$-fields 
were eliminated or not; here we give only results in the $F$-eliminated 
case, which means that in the interests of notational simplicity we here 
represent as unbarred quantities which appeared barred 
($\mbar^2, \hbar\cdots$) in Ref.~\jjwa. 

We begin with  a superpotential of the form:

\eqn\newW{
 W (\phi) = \frak{1}{2}{\mu}^{ij}\phi_i\phi_j + \frak{1}{6}Y^{ijk}
\phi_i\phi_j\phi_k,}
and soft \sy-breaking scalar terms as follows:
\eqn\newV{\eqalign{V_{\hbox{soft}} &= \left(c^i\phi_i 
+ \frak{1}{2}b^{ij}\phi_i\phi_j 
+ \frak{1}{6}h^{ijk}\phi_i\phi_j\phi_k +\hbox{c.c.}\right)
+(m^2)^i{}_j\phi_i\phi^j,\cr}} 
Note that $V_{\hbox{soft}}$ contains a linear term, so we are 
allowing for gauge singlets in general. As remarked in Ref.~\jjwa,
in the presence of soft breakings we can without loss of generality 
omit the linear term from the superpotential $W$. This statement follows 
(in the single field case) simply from the identity
\eqn\singrel{
V = |a + \mu\phi + y\phi^2/2|^2 = |\mu\phi + y\phi^2/2|^2 +  
(c\phi + b\phi^2/2 + h.c.) + a^* a,}
where $c = a^*\mu$ and $b = a^* y$, which shows how a linear term 
can be ``removed" from the superpotential. 
In the absence of explicit supersymmetry 
breaking this particular 
toy model has, of course \sic\ ground states, corresponding 
to $V = 0$ (i.e. it is not of the O'Raifeartaigh 
\ref\lian{L.~O'Raifeartaigh, \npb 96 (1975) 331} type). 

The complete exact results for the soft $\beta$-functions
are given by:
\eqn\allbetas{\eqalign{
\beta_M &= 2\Ocal\left[\frakk{\beta_g}{g}\right],\cr
\beta_{h}^{ijk} &= h{}^{l(jk}\gamma^{i)}{}_l -
2Y^{l(jk}\gamma_1{}^{i)}{}_l, \cr
\beta_{b}^{ij} &=
b{}^{l(i}\gamma^{j)}{}_l-2\mu{}^{l(i}\gamma_1{}^{j)}{}_l
+Y^{ijl}\sigma_l,\cr
\beta^i_{c} &= c^j \gamma^i{}_j
+\Delta Z^i+\mu^{il}\sigma_l-(m^2)^i{}_kZ^k,\cr
\left(\beta_{m^2}\right){}^i{}_j &= \Delta\gamma^i{}_j,\cr}}
where 
\eqna\Otdef$$\eqalignno{
{\cal O}  &= Mg^2\frakk{\partial}{\partial g^2}-h^{lmn}
\frakk{\partial}{\partial Y^{lmn}}-b^{lm}\frakk{\pa}{\pa \mu^{lm}},
&\Otdef a\cr
\Delta &= 2\Ocal\Ocal^* +2MM^* g^2{\partial
\over{\partial g^2}}
+\left[\Ytil^{lmn}{\partial\over{\partial Y^{lmn}}}
+\mutil^{lmn}{\partial\over{\partial \mu^{lm}}}+ \hbox{c.c.}\right]
+X{\partial\over{\partial g}},&\Otdef b\cr
(\gamma_1)^i{}_j  &= {\cal O}\gamma^i{}_j,
&\Otdef c}$$
\eqn\tydef{
\Ytil^{ijk} = (m^2)^{(i}{}_lY^{jk)l} \quad\hbox{and}\quad
\mutil^{ij} = (m^2)^{(i}{}_l\mu^{j)l},}
and $\sigma$, $Z^i$ are  defined as follows:
 \eqna\Zdef$$\eqalignno{
Z_i &= Y_{imn}K^{mn}{}_{pq}\mu^{pq}, &\Zdef a\cr
\sigma_i &= -2{\cal O}\left( Z_i\right)&\Zdef b,}$$
where $Y_{imn} = (Y^{imn})^*$, 
with $K^{mn}{}_{pq}$ defined by the condition
\eqn\Xdef{
Y_{imn}K^{mn}{}_{pq}Y^{pqj}a_j = \gamma^j{}_ia_j.}

Finally the $X$ function above is given (in the NSVZ scheme)
\eqn\exX{
X^{\NSVZ}=-2{g^3\over{16\pi^2}}
{S\over{\left[1-2g^2 C(G)(16\pi^2)^{-1}\right]}}}
where
\eqn\Awc{
S =  r^{-1}\tr [m^2C(R)] -MM^* C(G),}
For a discussion of $X$ in the $\DREDp$ scheme, see 
Ref.~\ref\jjpb{I.~Jack, D.R.T.~Jones and A.~Pickering,
\plb 432 (1998) 114}.

From Eq.~\Zdef{a}, \Xdef\ we see that 
$Z$ is obtained from the subset of contributions to the 
anomalous dimension $\gamma^j{}_i$ where the external lines are attached 
to Yukawa (as opposed to gauge) couplings, 
by replacing the ``outer" 
$Y^{jpq}$ by a supersymmetric mass insertion $\mu^{pq}$. 
Manifestly $Z$ is only nonzero 
if there are gauge-singlet chiral superfields.

The following set of relations are RG invariant to all orders of 
perturbation theory:

\eqna\amsball$$\eqalignno{
M &= M_0{\beta_g\over g},&\amsball a\cr
h^{ijk} &= -M_0\beta_Y^{ijk}, &\amsball b\cr
(m^2)^i{}_j &= \frak{1}{2}|M_0|^2\mu{d\gamma^i{}_j\over{d\mu}}, 
&\amsball c\cr
b^{ij} &= -M_0\beta_{\mu}^{ij} - M_0Y^{ijk}Z_k, &\amsball d\cr
c^i &= \frak{1}{2}|M_0|^2
\left[\mu{d\over{d\mu}}Z^i+(\gamma Z)^i\right]-M_0\mu^{il}Z_l. &\amsball d\cr
}$$
In this paper, in common with most of the AMSB literature,
we focus on Eqs.~\amsball{a-c}, by choosing theories such that $Z_i = 0$, and 
assuming an alternative source for $b^{ij}$, so that it can be treated 
as a free parameter and determined 
in the usual way by the minimisation conditions at the weak scale. 
Because the relevant $\beta$-functions do not involve $b^{ij}$ and $c^i$,
Eqs.~\amsball{a-c}\ form a RG-invariant subset.

For a $U_1$ theory we can also consider 
the possibility of a FI term, $\xi D$. 
In the unbroken theory $\xi$ is unrenormalised, as long as  
$\Tr \Ycal = 0$, where $\Ycal$  is the 
hypercharge matrix\ref\fisch{W.~Fischler et al, \prl 47 (1981) 757}. 
In the presence of soft breaking, the renormalisation of 
$\xi$ has been studied up to three loops
\ref\xius{I.~Jack and D.R.T.~Jones, \plb473 (2000) 
102}\nref\jjp{I.~Jack, D.R.T.~Jones and S.~Parsons, \prd62 (2000) 125022}%
--\ref\jjxib{I.~Jack and D.R.T.~Jones, \prd63 (2001) 075010}; 
however there exists no exact relation 
expressing its $\beta$-function in terms of $\beta_g$ 
and $\gamma$ of the type Eq.~\allbetas.

For a $U_1$ theory, the AMSB  solution for $m^2$ in the 
$D$-eliminated case may be generalised to either 
\eqn\massrga{
(m^2)^i{}_j = \frak{1}{2}|M_0|^2\mu{d\gamma^i{}_j\over{d\mu}}
+g\xi^{\rm RG}(\Ycal)^i{}_j,} 
where $\xi^{\rm RG}$ is the RG solution for $\xi$,
or 
\eqn\massrgb{
(m^2)^i{}_j = \frak{1}{2}|M_0|^2\mu{d\gamma^i{}_j\over{d\mu}}
+\zeta (\Ycal)^i{}_j,} where $\zeta$ is a constant.   
It is the latter possibility that forms Case~(FI) of this paper. 

If the theory admits a ${\cal R}$ symmetry then there is an alternative 
which is also exactly RG invariant,
\eqn\Rsola{
(m^2)^i{}_j = \frak{1}{2}|M_0|^2\mu{d\gamma^i{}_j\over{d\mu}} +
m_0^2 (\gamma^i{}_j + \qbar^i \delta^i{}_j),}
 where 
\eqn\qcharga{\qbar^i = 1 -\frak{3}{2}r^i,}
and $r^i$ are the ${\cal R}$-charges. This is Case~$\Rcal$ in this paper.

\bigskip\centerline{{\bf Acknowledgements}}

DRTJ was  supported by a PPARC Senior Fellowship and a Visiting 
Fellowship at All Souls College Oxford, and is grateful  to the Warden
and Fellows of All Souls, and the members of the Oxford  Theoretical
Physics Department, for hospitality.  We also thank Graham Ross and 
Stuart Raby   for helpful conversations.

\listrefs

\end